\documentclass[a4paper,10pt]{article}

 \usepackage[cp1251]{inputenc}

\usepackage{cite,amsmath,amsfonts,amsthm,fullpage}
\usepackage{youngtab}

\newcommand{\bpow}{\mathbf{p}}
\newcommand{\bbpow}{{\bar{\mathbf{p}}}}

\newcommand{\tU}{\textsc{U}}
\newcommand{\tZ}{\textsc{Z}}

\theoremstyle{plain}

\newtheorem{Lemma}{Lemma}
\newtheorem{Proposition}{Proposition}
\newtheorem{Corollary}{Corollary}
\newtheorem{Remark}{Remark}

\theoremstyle{remark}

\def\l{\langle}
\def\r{\rangle}

\def\Tr{\mathrm {Tr}}
\def\tr{\mathrm {tr}}
\def\det{\mathrm {det}}

\def\res{\mathop{\mathrm {res}}\limits_}

\def\bp{\begin{Proposition}}
\def\ep{\end{Proposition}}
\def\bc{\begin{Corollary}}
\def\ec{\end{Corollary}}
\def\bl{\begin{Lemma}}
\def\el{\end{Lemma}}
\def\be{\begin{equation}}
\def\ee{\end{equation}}
\def\br{\begin{Remark}\rm\small}
\def\er{\end{Remark}}
\def\brs{\begin{remarks}.\\ \rm\
\begin{enumerate}}
\def\ers{\end{enumerate}\end{remarks}}
\def\bea{\begin{eqnarray}}
\def\eea{\end{eqnarray}}

\def\Tr{\mathrm {Tr}}
\def\tr{\mathrm {tr}}
\def\det{\mathrm {det}}

\def\res{\mathop{\mathrm {res}}\limits}

\def\&{&{\hskip -20pt}}

\newcount\YDcount\YDcount=0
\def\YDsize{10pt}

\def\YD#1{%
\ifnum#1=0
 \ifnum\YDcount=0 \ifx\varnothing\undefined\emptyset\else\varnothing\fi
 \else\vskip1.4pt\egroup\YDcount=0\fi
\else
 \ifnum\YDcount=0 \YDcount=1\vcenter\bgroup\vskip1pt
 \else\nointerlineskip\fi
 \vbox{\hrule\hbox{\vrule height\YDsize
 \loop\hskip\YDsize\vrule\ifnum\YDcount<#1\advance\YDcount1\repeat}\hrule
 \kern-0.4pt}\expandafter\YD
\fi}

\usepackage[usenames,dvipsnames]{color}
\usepackage{ulem}

\begin{document}

\author{ Aleksander Yu.
Orlov\thanks{Institute of Oceanology, Nahimovskii Prospekt 36,
Moscow 117997, Russia, and National Research University Higher School of Economics,
International Laboratory of Representation
Theory and Mathematical Physics,
20 Myasnitskaya Ulitsa, Moscow 101000, Russia, email: orlovs@ocean.ru
}}
\title{Matrix integrals and Hurwitz numbers}

\maketitle

\begin{abstract}

We consider multi-matrix models which may be viewed as integrals of products of tau functions of matrix argument. 
Sometimes such integrals are tau functions themselves. 
We consider models which generate Hurwitz numbers $H^{\textsc{e},\textsc{f}}$, where
$\textsc{e}$ is the Euler characteristic of the base surface and $\textsc{f}$ is the number of branch points.
 We show that in case the integrands contains the product of $n > 2$ matrices 
 the integral generates Hurwitz numbers with Euler characteristic $\textsc{e}\le 2$ and the number of branch points $\textsc{f}\le n+2$,
 both numbers $\textsc{e}$ and $\textsc{f}$ depend on $n$ and the order of the multipliers in the matrix product. The number 
 $\textsc{e}$ may be even or odd (respectively describes Riemann (and certian Klein) or only Klein (non-orientable)
 base surface) depending on the presence of the BKP tau functions in the integrand.

\end{abstract}

\bigskip



$\,$

\begin{flushright}
 To the memory of Petya Kulish
\end{flushright}

\section{Introduction}

In the present paper we consider multi-matrix models where the interaction vertices contain products of $n$ interacting
matrices. Partition functions of such models are defined as integrals over entries of each matrix of a product of a certain 
number of tau functions where each tau function depends on products of matrices. This dependence of tau functions on
matrices is defined as follows. As it well known, a tau function of an integrable hierarchy depends on the set of higher times
$\{ p_m,\,m\ge 1\}$ of the hierarchy. In our case we put $p_m  =\tr Z^m$ where $Z$ denotes the product of $n$ matrices, then 
the tau function depends only on the eigenvalues of $Z$.
A tau function which depends on the eigenvalues of a matrix in such a way we call tau function of matrix argument.
In our paper we consider multi-matrix models which are integrals of  products of tau functions of matrix arguments.

The situation is well-known in case $n=1,2$, for instance, see \cite{O-2004-New} where integrals of products of tau functions
are considered and matrix integrals themself are tau functions and works 
\cite{GMMMO},\cite{KMMOZ},\cite{KMMM},\cite{Z},\cite{HO-2003},\cite{O-Acta},\cite{O-2004-New} where various matrix models used in physics
were considered. Integrals over unitary, complex, normal and Hermitian matrices were studied there.

What is needed in this paper is the character expansion
of matrix integrals, as was first presented in the works \cite{DiFIz}, \cite{Kazakov} and used in \cite{KMMM}, \cite{HO-2003}, \cite{O-2004-New} 
 in the context of relationships between matrix models and integrable systems. 
We note that in those cases when the matrix integral turned out to be a tau function, this one was either a tau function of the 
relativistic Toda lattice (commonly called ``TL tau function''), or the tau function of the BKP hierarchy,
introduced in the work of Kac and van de Leur \cite{KvdLbispec}.

 In this case, the perturbation theory series in the coupling constants turns out to be
 the so-called tau function of the hypergeometric type. For the hierarchy of the relativistic Toda lattice, a tau function of this type
 was written in \cite{KMMM} and was thoroughly studied in \cite{OS-2000}, \cite{OS-TMP}.
 
 We recall that the hierarchy of TL arises in one- and two-matrix models with hermitian and unitary matrices, and also in some others,
 in particular, in a number of matrix models described in the paper \cite{O-2004-New} (and in the Appendix to \cite{HO-2006}),
 and the hierarchy of BKP arises in the description of the orthogonal and symplectic ensembles \cite{L1},
  $ \beta = 1,4 $ of circular ensembles \cite{OST-II} and $ \beta = 1,4 $ Ginibre ensembles
 \cite{O-2012}, in the description of the Panda-Mehta ensembles interpolating between unitary and orthogonal or between unitary
 and symplectic ensembles \cite{OST-II}, and also in the description of two-matrix models with one Hermitian, one symmetric
 (Or anti-symmetric)  or with one Hermitian, the second self-dual (or anti-self-dual) matrices.

 The connections between various matrix integrals and certain types of Hurwitz numbers have been presented in many works
   \cite{MelloKochRamgoolam}, \cite{AMMN-2014}, \cite{Zog}, \cite{GGPN},
\cite{Chekhov-2014}, \cite{NO-2014}, \cite{NO-LMP}, \cite{Chekhov-2016}.
Recall that the Hurwitz numbers $ H^{\textsc{e}, \textsc{f}}(d; \Delta^1, \dots, \Delta^{\textsc{f}}) $
count the nonequivalent branched $ d $-fold coverings of the base Riemann or Klein surface with
Euler characteristic $ \textsc{e} $ by another Riemann or Klein surface,
with given $ \textsc{f} $ branching profiles $ \Delta^1, \dots, \Delta^{\textsc{f}} $. In the works cited
two cases were studied:

(1) covering of the Riemann sphere ($\textsc{e}=2$) such that the two profiles, $\Delta^1,\Delta^2$, are arbitrary,
and the lengths of all other profiles
are fixed, this case relates to the TL hierarchy of integrable equations \cite{JM}, \cite{UT} (or, equivalently, to
 the two-component KP hierarchy)

(2) covering of the (real) projective plane ($\textsc{e}=1$) with one arbitrary profile $\Delta^1$ and with fixed lengths 
of profiles at all other branch points. 
This case relates to the BKP hierarchy of V. Kac and J. van de Leur \cite{KvdLbispec}.

In the present paper we need more complicated matrix integrals in which more matrices participate, as in the works
\cite{Alexandrov}, \cite{AMMN-2014},\cite{ChekhovAmbjorn},\cite{NO-2014},\cite{NO-LMP}. 
Here I develop the proposed methods of the last two papers and of the work \cite{O-2004-New}.

Our goal is to present integrals \textit{of} tau functions and/or products of tau functions  which generate Hurwitz numbers
$H^{\textsc{e},\textsc{f}}$ with an arbitrary chosen Euler characteristic of the base surface $\textsc{e}$ and an arbitrary number
of branch points $\textsc{f}$. 
We start with matrix integrals which are tau functions themselves. Such integrals-tau functions always generate Hurwitz numbers either for the
Riemann sphere ($\textsc{e}=2$) or for the real projective plane ($\textsc{e}=1$). Let us note that
these integrals can be related \cite{OrlovStrahov} to the popular problem of quantum chaos: the study of eigenvalues of the product of 
random matrices \cite{Ak1}, \cite{Ak2}, \cite{AkStrahov}, \cite{S1}, \cite{S2}.
Next, we show that a change of the order in the products of matrices yields different generating functions for Hurwitz numbers
where the Euler characteristic of the base surface $\textsc{e}$ is reduced by an even number. 
(Technically this phenomenon is caused by Remarks \ref{Deg} and \ref{Euler=Deg}). In this case the matrix integral 
is not a tau function. 
(Nevertheless, such integrals inherit some properties of tau functions.)
The present work is a continuation of the work \cite{O-2004-New} in the following sense: in the work \cite{O-2004-New} only 
such integrals of TL tau
functions (or products of tau-functions), which in turn could be interpreted as tau functions, but in the present
there is no such restriction. In addition to the tau functions of the Toda lattice, we also admit the BKP tau functions under the integral 
(the latter give odd Euler characteristics, which correspond to non-orientable surfaces).

The paper consists of the Introduction, two sections and Appendices.  In the introductionary subsections below, we introduce  
the objects we need and the main technical Lemma \ref{useful-relations}, which was used in the previous work \cite{O-2004-New}.
We also define special combinations of Hurwitz numbers (\ref{GJ-Hurwitz-number'}) which
generalise the so-called monotone Hurwitz numbers introduced by Goulden and Jackson in the well-known paper
 \cite{Goulden-Jackson-2008}. Let us mark that this includes double Hurwitz numbers, introduced earlier
 in the pioneering work \cite{Okounkov-2000} of Okounkov; the Hurwitz numbers related to the so-called completed cycles from 
 \cite{Okounkov-Pand-2006}  
 can be also expressed via Goulden-Jackson ones (this also applies to the "integrable" case in the work of Mironov, Morozov and 
 Natanzon \cite{MM1}). 
 We define a generalised hypergeometric function which generates such combinations for arbitrary $\textsc{e}$, see (\ref{gen-hyp-funct}).
Section \ref{Complex} is devoted to the product of complex matrices. Section
\ref{Unitary} is almost identical to the previous one, but unitary matrices are replaced by complex ones.
The Appendices are devoted to a survey of necessary information on the concepts of partitioning, Schur functions, Hurwitz numbers, 
and we also recall some facts on integrable systems and tau functions. We repeat some parts of the works \cite{NO-2014} and \cite{NO-LMP}.
In particular, we discuss how to calculate the Hurwitz numbers for Riemann and Klein surfaces
with Euler characteristics of the covering surface smaller than $ \textsc{e} = 1 $, see Lemma \ref{Hurwitz-down-Lemma}
and the Proposition \ref{d-cycle-proposition}. I cite matrix integrals taken from \cite{NO-2014} and \cite{NO-LMP} which
generate similar Hurwitz numbers, but they differ from those given in the main text of this paper.

Once in Galipolli meeting (I guess it was in 2002) I gave a talk called ``Hypergeometric functions, symmetric function and matrix integrals''.
We discussed and Petya Kulish remarked that it would be intresting to use the characters of the symmetric group 
 in applications of the soliton theory. I am grateful to him for the discussion. This work to some extent meets his wish.

\subsection{Product of complex matrices\label{Product-of-complex-matrices}}

Let us consider a set of $N\times N$ complex matrices $Z_\alpha,\,C_\alpha,\,\,\alpha=1,\dots,n$.
Here and below the label ``star'' does not denote the complex conjugate.
And define the following products
\bea\label{Z}
\tZ&:=&(Z_1 C_1) \cdots (Z_n C_n)\\ 
\label{Z^*}
\tZ^*&:=& Z_n^\dag\cdots Z_1^\dag\\
\label{tildeZ^*}
\tZ^{(t)}&:=& Z^\dag_n Z^\dag_{n-1} \cdots Z^\dag_{t+1} Z^\dag_1Z^\dag_2  \cdots Z^\dag_{t} ,\qquad t \le n
\eea
where $Z_\alpha^\dag$ is the Hermitian conjugate of $Z_\alpha$. (The matrix $\tZ^{(t)}$ may be obtained 
by the action of the product of $[\frac t2]$ transpositions on $\tZ^*$).
In the formula (\ref{tildeZ^*}) in the case $ t = n $ we assume
$ \tZ^{(n)} = Z^\dag_1Z^\dag_2 \cdots Z^\dag_n $. We have $ \tZ^{(0)} = \tZ^* $.

We also introduce notations
\be\label{tau_1}
\tau_1^{\rm TL}(X,\bpow)=\sum_\lambda s_\lambda(X)s_\lambda(\bpow) =e^{\tr V(X,\bpow)}=
\prod_{i=1}^N e^{\sum_{m=1}^\infty \frac{1}{m}x_i^mp_m}
\ee
where $x_i$ are eigenvalues of $X$,
where $\bpow=(p_1,p_2,\dots)$ is a semi-infinite set of parameters,
and
\be\label{tau_1^B}
 \tau_1^{\rm BKP}(X)=\sum_\lambda s_\lambda(X) =\prod_{i=1}^N (1-x_i)^{-1}\prod_{i<j} (1-x_ix_j)^{-1}
\ee
Here $s_\lambda$ denotes the Schur function, see Section \ref{Partitions-and-Schur-functions} in Appendix. We recall the fact
\cite{Mac} we shall need: if $X$ is $N\times N$ matrix, then
\be\label{ell<N}
s_\lambda(X)=0,\qquad {\rm if}\quad \ell(\lambda)>N
\ee
where $\ell(\lambda)$ is the length of a partition $\lambda=(\lambda_1,\dots,\lambda_\ell),\,\lambda_\ell >0$.

We will study integrals over complex matrices where the measure is defined as
\be
d\mu(Z_\alpha)=c \prod_{i,j=1}^N d\Re (Z_\alpha)_{ij}d\Im (Z_\alpha)_{ij}e^{-|(Z_\alpha)_{ij}|^2}
\ee
where the integration range is $\mathbb{C}^{N^2}$ and where $c$ is the normalization constant defined via $\int d\mu(Z_\alpha)=1$.

We treat this measure as probability measure. The related ensemble is called the ensemble of random complex matrices or, the same, the complex Ginibre enesemble. 
The expectation of a quantity
$f$ which depends on entries of a matrix $Z$ is defined by
$$
\mathbb{E}(f)=\int f(Z) d\mu(Z).
$$

We also consider integrals over the unitary group $\mathbb{U}(N)$, $d_*U$ denotes the related Haar measure 
(see (\ref{Haar-unitary}) in the Appendix), $\int_{\mathbb{U}(N)} d_*U=1$.

To evaluate integrals we apply relations used in \cite{O-2004-New} 
and \cite{NO-2014}, \cite{NO-LMP}
(for instance see \cite{Mac} for the derivation). 
\bl \label{useful-relations}
Let $A$ and $B$ be normal  matrices (i.e. matrices diagonalizable by unitary transformations). Then
\begin{equation}\label{sAUBU^+1}
\int_{\mathbb{U}(N)}s_\lambda(AUBU^{-1})d_*U=
\frac{s_\lambda(A)s_\lambda(B)}{s_\lambda(\mathbb{I}_N)} \ ,
\end{equation}
For $A,B\in GL(N)$ we have
\begin{equation}\label{sAUU^+B'}
\int_{\mathbb{U}(n)}s_\mu(AU)s_\lambda(U^{-1}B)d_*U=
\frac{s_\lambda(AB)}{s_\lambda(\mathbb{I}_N)}\delta_{\mu,\lambda}\,.
\end{equation}
Below $p_{\infty}=(1,0,0,\dots)$. 
\begin{equation}\label{sAZBZ^+'}
\int_{\mathbb{C}^{n^2}} s_\lambda(AZBZ^+)e^{-\textrm{Tr}
ZZ^+}\prod_{i,j=1}^n d^2Z=
\frac{s_\lambda(A)s_\lambda(B)}{s_\lambda(p_{\infty})}
\end{equation}
and
\begin{equation}\label{sAZZ^+B'}
\int_{\mathbb{C}^{n^2}} s_\mu(AZ)s_\lambda(Z^+B) e^{-\textrm{Tr}
ZZ^+}\prod_{i,j=1}^nd^2Z= \frac{s_\lambda(AB)}{s_\lambda(p_{\infty})}\delta_{\mu,\lambda}\,.
\end{equation}
\el
This Lemma is a tool to study integrals of tau functions: the simplest $\tau_1^{\rm TL}$, $\tau_1^{\rm BKP}$ and more general series
(\ref{hyperg2KP-sums}) and (\ref{hypergBKP-sums}) below.
\br\label{Deg}
Notice that if we asign the degree 1 to each Schur function: ${\rm Deg}\, s_\lambda=1$  then relations 
(\ref{sAUBU^+1}) and (\ref{sAZBZ^+'}) keep
the degree while the degree of the right hand sides of both (\ref{sAUU^+B'}) and (\ref{sAZZ^+B'}) less that the degree of 
the left hand side by 2.
\er

\subsection{Schur functions, specializations and characteristic map relation}

We send the reader to Section  \ref{Partitions-and-Schur-functions} in Appendix for definitions and notations in particular, related to 
the notion of a partition.
Let us write down the Schur function as the quasi-homogenious polynomial in the so-called power sum variables $\bpow=(p_1,p_2,\dots )$ 
\cite{Mac}:
\be\label{Schur-char-map}
s_\lambda(\bpow)=\frac{{\rm dim}\lambda}{d!}\left(p_1^d+\sum_{\Delta\atop |\Delta|=d } \varphi_\lambda(\Delta)\bpow_{\Delta}\right)
\ee
where $\bpow_\Delta=p_{\Delta_1}\cdots p_{\Delta_{\ell}}$ and where $\Delta=(\Delta_1,\dots,\Delta_\ell)$ is a partition whose weight
coinsides with the weight of $\lambda$: $|\lambda|=|\Delta|$. Here 
\be
{\rm dim}\lambda =d!s_\lambda(\bpow_\infty),\qquad \bpow_\infty = (1,0,0,\dots)
\ee
is the dimension of the irreducable representation of the symmetric group $S_d$. We imply that 
$\varphi_\lambda(\Delta)=0$ if $|\Delta|\neq |\lambda|$.

Relation (\ref{Schur-char-map}) is known as the characteristic map relation (see for instance \cite{Mac}),
 it relates the Schur functions (irreducable characters of linear groups labeled by $\lambda$) 
to the irreducable characters  $\chi_\lambda$  of the symmetric group $S_d$, see (\ref{varphi}) below.

Given a matrix $A$, we will use the notation $s_\lambda(A):=s_\lambda(\bpow(A)$, where 
$\bpow(A)=\left(p_1(A),p_2(A),\dots\right)$ and
$p_k(A)= \tr A^k$. Say, for the unity $N\times N$ matrix $\mathbb{I}_N$, we have $p_k(\mathbb{I}_N)=N$ for 
any $k$. 

We shall use specializations of the Schur functions given by
\bl\label{specializations} Denote 
\be\label{p_infty}
\bpow_\infty =(1,0,0,\dots)
\ee
\be\label{p(a)}
\bpow(a)=\left(a,a,a,\dots \right)
\ee
\be\label{p(t,q)}
\bpow(\texttt{q},\texttt{t})=\left(p_1(\texttt{q},\texttt{t}),p_2(\texttt{q},\texttt{t}),\dots\right)\,,\quad
p_m(\texttt{q},\texttt{t})=  \frac{1-\texttt{q}^m}{1-\texttt{t}^m}
\ee 
Then
\be\label{Schur-t(a)}
\frac{s_\lambda(\bpow(a))}{s_\lambda(\bpow_\infty)}=(a)_\lambda\,,\quad \bpow(a)=(a,a,a,\dots)
\ee
where $(a)_\lambda:=(a)_{\lambda_1}(a-1)_{\lambda_2}\cdots (a-\ell+1)_{\lambda_\ell}$, $(a)_n:=a(a+1)\cdots(a+n-1)$,
where $\lambda=(\lambda_1,\dots,\lambda_\ell)$ is a partition.
More generally
\be\label{Schur-t(t,q)}
\frac{s_\lambda(\bpow(\texttt{q},\texttt{t}))}{s_\lambda(\bpow(0,\texttt{t}))}=(\texttt{q};\texttt{t})_\lambda\,,
\ee
where $(\texttt{q};\texttt{t})_\lambda =
(\texttt{q};\texttt{t})_{\lambda_1}(\texttt{q t}^{-1};\texttt{t})_{\lambda_2}\cdots
(\texttt{q t}^{1-\ell};\texttt{t})_{\lambda_\ell}$ where
$(\texttt{q};\texttt{t})_k=(1-\texttt{q})(1-\texttt{q t})\cdots (1-\texttt{q t}^{n-1}) $
is $\texttt{t}$-deformed Pochhammer symbol. $(\texttt{q};\texttt{t})_0=1$ is implied.
\el
This Lemma may be easily derived from known relations written down in \cite{Mac}.

Since both Pochhammer symbols (\ref{Schur-t(a)}) and (\ref{Schur-t(t,q)}) are equal to the ratio of Schur
functions, we assign ${\rm Deg}\,(a)_\lambda =0$ and ${\rm Deg}\,(\texttt{q};\texttt{t})_\lambda =0$.

\br
Let us notice that in the notations which we use we have
\be\label{unit-matrix}
s_\lambda(\mathbb{I}_N)=s_\lambda(\bpow(N))
\ee
\er

Let us consider
the sums of all normalized characters $\varphi_\lambda$  evaluated on partitions $\Delta$ with a given weight
$d$, $d=|\lambda|=|\Delta|$ and a given length $\ell(\Delta)=d-k$:
\be\label{phi}
\phi_k(\lambda)\,:=\,\sum_{\Delta\atop \ell(\Delta)=d-k} \,\varphi_\lambda(\Delta),\quad k=0,\dots,d-1
\ee
\br \label{special-cases}
Let us note that $\phi_0(\lambda)=1$.
There are two other special cases when the sum of normalized characters (\ref{phi}) contains a single term:

(a) $\phi_{1}(\lambda)=\varphi_\lambda(\Gamma),\, \Gamma=(1^{d-2}2)$ (for $d>1$). Notice that
$\phi_1(\lambda)=\varphi_\lambda(\Gamma)$ which is related to the \textit{ minimally ramified} profile:
the profile with the colength equal to 1.
This is the profile of the simple branch point, simple branch points are of main interest in many applications \cite{Dijkgraaf}.

(b) $\phi_{d-1}(\lambda)=\varphi_\lambda((d))$ which is related to the cyclic profile which describes the
\textit{ maximally ramified}  profile (this profile plays a specific role, see Proposition \ref{d-cycle-proposition}).

In what follows we shall use sums $\phi_k$ as building blocks to construct weighted sums of the Hurwitz numbers
(see for instance (\ref{GJ-Hurwitz-number}) and (\ref{tilde-phi}),(\ref{gen-Hurwitz-number}) below). 
Then the cases (a),(b) produce not the weighted sums but Hurwitz numbers themselves (see
(\ref{tau-Hurwitz-themselves}) below).
\er

\br\label{Hurwitz-genus-formula} The quantity $d-\ell(\lambda)$ which is used in the definition (\ref{phi}) is called the
\textit{colength} of a partition $\lambda$ and will be denoted by $\ell^*(\lambda)$.
The colength enters the so-called Riemann-Hurwitz formula which relates the Euler characteristic of a base surface,
$\textsc{e}$, to the Euler characteristic of it's $d$-branched cover,  $\textsc{e}'$ as follows
\[
 \textsc{e}'- d\textsc{e}+\sum_{i} \ell^*(\Delta^{(i)})=0
\]
where the sum ranges over all branch points $z_i\,,i=1,2,\dots$ with ramification profiles given by partitions $\Delta^i\,,i=1,2,\dots$
respectively.
\er

Let us introduce
\be\label{degree}
{\rm deg} \, \phi_k(\lambda)= k
\ee
This degree is equal to the colength of ramification profiles in formula (\ref{phi}), and due to Remark 
\ref{Hurwitz-genus-formula}
it will be important later to define the Euler characteristic of the covering surfaces in the parametrization I cases.

   We need
\be
s_\lambda(\bpow(a))=
\frac{{\rm dim}\lambda}{d!}a^d \left(1+\sum_{d > k > 0 } \phi_{k}(\lambda)a^{-k}\right)\,,\quad d=|\lambda|
\ee
which is the combination of (\ref{Schur-char-map}) and (\ref{Schur-t(a)}),
and its consequence
\be
\left( s_\lambda(\bpow(a))\right)^c =\left(\frac{{\rm dim}\lambda}{d!}\right)^c a^{cd}
\left(1+\sum_{m>0}\left(\sum_{d>  k > 0 } \phi_{k}(\lambda)a^{-k}\right)^m\right)^c=
\ee
\[
 =:\left(s_\lambda(\bpow_\infty) \right)^c a^{cd}\left(1+\sum_{k >0}{\tilde\phi}_k(\lambda;c) a^{-k}   \right)
\]
where each ${\tilde\phi}_k$ is built of the collection $\{ \phi_i\,,\,i>0 \}$ as follows:
\be\label{tilde-phi}
{\tilde\phi}_k(\lambda;c)=\sum_{l\ge 1}\, c(c-1)\cdots (c-l+1)\,
\sum_{\mu\atop \ell(\mu)=l,\,|\mu|=k}\frac{\phi_\mu(\lambda)}{|{\rm Aut}\,\mu |}\,,\quad \phi_\mu(\lambda):=\phi_{\mu_1}(\lambda)\cdots \phi_{\mu_l}(\lambda) 
\ee
where $\mu=(\mu_1,\dots,\mu_{l'})$ is a partition which may be written alternatively \cite{Mac} as 
$\mu=\left( 1^{m_1}2^{m_2}3^{m_3} \cdots\right)$
where $m_i$ is the number of times a number $i$ occurs in the partition of $|\mu|=k$. Thus the set  of all non-vanishing 
$m_{j_a}\,,a=1,\dots l'$, ($l'\le l$)
defines the partition $\mu$ of length $\ell(\mu)=\sum_{a=1}^{l'} m_{j_a} = l$ and of weight $|\mu|=\sum_{a=1}^{l'} j_a m_{j_a} =k$. Then
the order of the automorphism group of the partition $\mu$ is
\[
  |{\rm Aut}\,\mu | = m_{j_1}!\cdots m_{j_{l'}}!
\]

As we see ${\tilde\phi}_k(\lambda;1)={\phi}_k(\lambda)$.
We have
\be
{\rm deg}\,{\tilde \phi}_k =k
\ee

\subsection{Hurwitz numbers.} 

For all necessary details we need to understand relations of this subsection we send the reader to the paper \cite{NO-LMP}, 
the related parts of the mentioned work may be found in Section \ref{Hurwitz-numbers} in Appendix.

Introduce
\be\label{Hurwitz-number}
H^{\textsc{e},k}_N(d;\Delta^1,\dots,\Delta^{k})=\sum_{\lambda\atop |\lambda|=d,\,\ell(\lambda)\le N}
\left(\frac{{\rm dim}\lambda}{d!} \right)^{\textsc{e}}\varphi_\lambda(\Delta^1)\cdots 
\varphi_\lambda(\Delta^k)
\ee
where $\Delta^i$ is a partition labeled by $i=1,\dots,k$.
${\rm dim}\lambda$ is the dimension of the irreducible representation of $S_d$, and
\be
\label{varphi}
\varphi_\lambda(\Delta^{(i)}) := |C_{\Delta^{(i)}}|\,\,\frac{\chi_\lambda(\Delta^{(i)})}{{\rm dim}\lambda} ,
\quad {\rm dim}\lambda:=\chi_\lambda\left((1^d)\right)
\ee
$\chi_\lambda(\Delta)$ is the character of the symmetric group $S_d$ evaluated at a cycle type $\Delta$,
and $\chi_\lambda$ ranges over the irreducible complex characters of $S_d$ (they are
labeled by partitions $\lambda=(\lambda_1,\dots,\lambda_{\ell})$ of a given weight $d=|\lambda|$). It 
is supposed that $d=|\lambda|=|\Delta^{1}|=\cdots =|\Delta^{k}|$.
$|C_\Delta |$ is the cardinality of the cycle
class $C_\Delta$ in $S_d$, see (\ref{C-Delta,z-Delta}) in Appendix.

The character formula by Mednykh-Pozdnyakova-Jones \cite{M2},\cite{GARETH.A.JONES} (see Appendix \ref{Hurwitz-numbers} ) 
says that
for $N\ge d$ the number $H^{\textsc{e},k}_N(d;\Delta^1,\dots,\Delta^{k})$ plays the role of the Hurwitz number which counts
$d$-fold branched covers of a (base) Klein surface of the Euler characteristic $\textsc{e}$ by Klein surfaces of Euler characteristics
equal to $\textsc{e}'=d\textsc{e}+\sum_{i=1}^k(\ell(\Delta^i)-d)$.

\br\label{Euler=Deg}
Notice that if we pick up Hurwitz numbers from a ratio of the Schur functions as it was done in many papers 
(see, for instance \cite{NO-2014}) then by formula (\ref{Hurwitz-number}) the degree ${\rm Deg }$ 
(defined in Remark \ref{Deg})
yields the Euler characteristic $\textsc{e}$ of the base surface.

\er

\paragraph{Goulden-Jackson Hurwitz numbers.}

Let us introduce the following sum of Hurwitz numbers:
\be\label{GJ-Hurwitz-number}
S^{\textsc{e},k+p}(N;d;\Delta^1,\dots,\Delta^k;l_1^*,\dots,l_p^*):=
\sum_{\lambda\atop |\lambda|=d,\,\ell(\lambda)\le N}
\left(\frac{{\rm dim}\lambda}{d!} \right)^{\textsc{e}} \prod_{i=1}^k\varphi_\lambda(\Delta^i)\prod_{i=1}^p 
\phi_{l_i^*}(\lambda)
\ee
\be\label{GJ-Hurwitz-number'}
=\sum_{\Delta^{k+1}\atop \ell^*\left(\Delta^{k+1}\right)=l_1^*} 
\sum_{\Delta^{k+p}\atop \ell^*\left(\Delta^{k+p}\right)=l_p^*}\left(d;\Delta^{1},\dots,\Delta^{k+p}  \right)
\ee
where all weights $|\Delta^i|,\,i=1,\dots,k+p$ are considered to be the same and equal to $d=|\lambda|$.
The last expression is the sum of Hurwitz numbers related to the Euler characterisitic of the base equal to 
$\textsc{e}$ and the Euler characterisitic of $d$-fold branched covers equal to
\[
 \textsc{e}'=\textsc{e}d -\sum_{i=1}^k \ell^*(\Delta^i)-\sum_{i=1}^p l_i^*
\]
The number of branch points is $k+p$ where in $k$ branch points the ramification profiles are given by partitions
$\Delta^i,\,i=1,\dots,k$ and in the rest $p$ branch points the length of partitions are given respectively by 
$d-l_i^*,\,i=1,\dots,p$.

In the cases further considered in Sections \ref{Complex} and \ref{Unitary}, the parameter $N$ denotes the matrix size, 
we are free to send it to infinity, and if it not produces a confussion we shall omit the label $N$ in $H_N^{\textsc{e},k}$.

It was shown by Goulden and Jackson in \cite{Goulden-Jackson-2008} that the following generating function
\be
\sum_{\Delta}\sum_{l_1^*,\dots,l_p^*=1}^{d}\, S^{2,p+1}\left(d;\Delta;l_1^*,\dots,l_p^*  \right)\bpow_\Delta 
\prod_{i=1}^pa_i^{d-l_i^*}
\ee
is an example of a KP tau function, namely, the hypergeometric tau function of KP \cite{OS-2000}:
\be\label{hyp-tau}
\tau^{\rm TL}_r(n,\bpow,\bpow^*) = \sum_{\lambda} s_\lambda(\bpow)s_\lambda(\bpow^*)r_\lambda(n),\quad
r_\lambda(n):=
\prod_{(i,j)\in\lambda}r(n+j-i)
\ee
with the specification $\bpow^*=\bpow_\infty$ and $r(x)=\prod_{i=1}^p(x+a_i)$ where $a_i$ are arbitrary complex parameters.
The product in the right hand side of (\ref{hyp-tau}) is called content product. The important role of content products
in many of combinatorial problems related to the study of symmetric groups and Hurwitz numbers was presented in 
\cite{Goulden-Jackson-2008} (see references there and also \cite{Harnad-overview-2015}). 

In \cite{AMMN-2014} and in \cite{Harnad-2014} the Goulden-Jackson Hurwitz numbers $S^{2,p+2}$ related
to TL hypergeometric tau functions \cite{KMMM},\cite{OS-2000} were studied. 
It was shown in \cite{AMMN-2014} how these numbers may be generated by a multimatrix model different from
written down in the present paper.

Let us note that the projective analogues of Goulden-Jackson Hurwitz numbers $S^{1,p+1}$, namely, Goulden-Jackson Hurwitz numbers
where the base surface was the projective plane were introduced in \cite{NO-2014}, \cite{NO-LMP}. It was shown that projective 
Goulden-Jackson Hurwitz numbers are generated by a BKP hypergeometric tau function \cite{OST-I}, namely by
\be\label{hyp-tau-B}
\tau^{\rm B}_r(N,n,\bpow)=\sum_{\lambda\atop \ell(\lambda)\le N} s_\lambda(\bpow) r_\lambda(n)
\ee
with the same specialization of $r$ as was prevousely given for the hypergoemetric tau function of the KP hiererchy.
Here $N,n,\bpow$ is the set of free parameters called BKP higher times. The examples of content products
are Poschhammer symbol  attached to a Young diagram $\lambda$ and also and its $\texttt{t}$-deformed versions
(see Lemma \ref{specializations}):
\be\label{Pohh}
 (a)_\lambda=\prod_{(i,j)\in\lambda}\left(a+j-i \right)=\prod_{i=1}^{\ell(\lambda)}(a-i+1)_{\lambda_i}
 \,,\quad
 (\texttt{q};\texttt{t})_\lambda=\prod_{(i,j)\in\lambda}\left(1-\texttt{q}\texttt{t}^{j-i} \right)
\ee
where $(a)_k :=a(a+1)\cdots (a+k-1)$.

For future purposes we also write down the following specialization of the content product which may be expressed in terms of
the Schur functions 
\be
\prod_{i=1}^k \left( (a_i+N)_\lambda \right)^{n_i}
\prod_{i=1}^{k'} 
\left( (\texttt{q}_i\texttt{t}^{N-1};\texttt{t}_i)_\lambda \right)^{{\tilde n}_i} 
 =\prod_{i=1}^k 
 \left( \frac{s_\lambda(\bpow(a_i+N))}{s_\lambda(\bpow_\infty)} \right)^{n_i}
 \prod_{i=1}^{k'} 
\left(\frac{s_\lambda(\bpow(\texttt{q}_i\texttt{t}^{N-1};\texttt{t}_i))}{s_\lambda(\bpow(0;\texttt{t}_i))} \right)^{{\tilde n}_i}
\ee
For such content products both hypergeomentric tau functions (\ref{hyp-tau}) and (\ref{hyp-tau-B}) are expressed entirely in terms of 
the Schur functions,
and, therefore, in terms of Hurwitz numbers and the generating parameters, see \cite{HO-2014},\cite{NO-2014},\cite{NO-LMP}.

\br\label{Deg-tau}
In \cite{NO-LMP} it was shown that not only Pochhammer symbols but rather general content products may be expressed in terms of
ratios of Schur functions and therefore ${\rm Deg}\,\tau_r^{\rm TL} =2$ and ${\rm Deg}\,\tau_r^{\rm BKP} =1$.
\er

\paragraph{Generalized Hurwitz numbers. Hypergeometric series.}

Let us introduce the weighted sums of Hurwitz numbers as follows:
\be\label{gen-Hurwitz-number}
G^{\textsc{e},k;p}\left(N;d;\Delta^1,\dots,\Delta^k {|\atop |} {k_1,\dots,k_p \atop n_1,\dots, n_p}\right):=
\ee
\be
\sum_{\lambda\atop |\lambda|=d,\,\ell(\lambda)\le N}
\left(\frac{{\rm dim}\lambda}{d!} \right)^{\textsc{e}} \prod_{i=1}^k\varphi_\lambda(\Delta^i)\prod_{i=1}^p
{\tilde\phi}_{k_i}(\lambda;n_i)
\ee
For large enough $N$ (namely $N\ge \ell(\lambda)$) this is a weighted sum of Hurwitz numbers which count non-equivalent $d$-fold coverings
of the connected Klein surface with the Euler characteristic $\textsc{e}$ (the base surface) by (not necessarily connected) Klein
surfaces with the Euler characteristic 
\[
 \textsc{e}'= d\textsc{e}-\sum_{i} \ell^*(\Delta^{(i)})-\sum_{j=1}^p k_j
\]
As it follows from (\ref{tilde-phi}) the weights mentioned above depends on the set $\{ n_j\}$.

The generalised Hurwitz numbers are generated by the following hypergeometric series in $k$ variables $\bpow^{(i)}\,,\,i=1,\dots,k$:
\be\label{gen-hyp-funct}
F^{\textsc{e},k;p}\left({a_1,\dots,a_p \atop n_1,\dots, n_p}{|\atop |}\bpow^{(1)},\dots,\bpow^{(k)}  \right)=
\sum_{\lambda\atop\ell(\lambda)\le N}
\, \left(s_\lambda(\bpow_\infty)\right)^{\textsc{e}-\sum_{j=1}^p n_j -k}\,
\prod_{j=1}^p \left( s_\lambda(\bpow(a_j))\right)^{n_j} \,
\prod_{i=1}^k s_\lambda(\bpow^{(i)})
\ee
Here we consider $a_i\,,\,i=1,\dots,p$ to be independent parameters.

The right hand side of (\ref{gen-hyp-funct}) may be rewritten with the help of Pochhammer symbols:
 \be\label{gen-hyp-funct-Pochhammer}
 F^{\textsc{e},k;p}\left({a_1,\dots,a_p \atop n_1,\dots, n_p}{|\atop |}\bpow^{(1)},\dots,\bpow^{(k)}  \right)
   = \sum_{\lambda\atop\ell(\lambda)\le N}
\, \left(\frac{{\rm dim}\lambda}{|\lambda|!}\right)^{\textsc{e}-\sum_{j=1}^p n_j -k}\,
\prod_{j=1}^p \left((a_j)_\lambda\right)^{n_j}  \,
\prod_{i=1}^k s_\lambda(\bpow^{(i)})
 \ee
Let us notice that according to Remark \ref{Deg} we get
\be
{\rm Deg}\,F^{\textsc{e},k;p}=\textsc{e}
\ee

 \br

 The introduced hypergeometric function can be considered as a version of some discrete $ \beta $-ensemble, since
 the sum over the partitions can be written as the sum over the cone $ h_1> \cdots> h_N \ge 0 $, 
 $ h_m = \lambda_m-m + N, \, m = 1, \dots, N $, and then 
 continue the summand to all pairwise non-coinciding non-negative
 values of $h_m$ using  the antisymmetry of the Schur function as a function of the variables $ h_m, \, m = 1, \dots, N $ 
 (see the definition  (\ref{Schur-t}) in the Appendix), and replace the sum over the cone 
 by a sum over all   pairwise non-coinciding non-negative values of $h_m$ 
 earning the factor $1/N!$.  We get 
\[
 F^{\textsc{e},k;p}_N\left({a_1,\ldots,a_p}{n_1,\ldots, n_p}\,\bigg|\,\mathbf p^{(1)},\ldots,\mathbf p^{(k)}\right)=
\]
\[
 =\frac{1}{N!}\sum_{h_1,\ldots,h_N}\rule{-10pt}{10pt}'\;\;\prod_{a<b}|h_a-h_b|^{\beta}\prod_{m=1}^N(h_m!)^{-\beta}e^{-V_{h_m}-U_{h_m}}
 \prod_{i=2}^k \Theta_{\{h\}} s_{\{h\}}(\mathbf p^{(i)})
\]
where $\beta=-\textsc{e}+\sum_{j=1}^p n_j +k$, and $\sum'$ denotes the summation over all different non-negative integers $h_i$, $i=1,\ldots,N$.
Here the following notations are introduced
\be
 e^{-V_{h_m}}=\prod_{j=1}^p((a_j-m+1)_{\lambda_m})^{n_j},\qquad
 e^{-U_{h_m}}=s_{(\lambda_m)}(\mathbf p^{(1)})\qquad \Theta_{\{h\}}=sign \prod_{a<b}(h_a-h_b) 
\ee
We shall omit the subscript $N$ below.
Next, let
\be
 p^{(i)}_j=\tr C_i^j=\sum_{m=1}^N e^{jy_m^{(i)}},\qquad i=1,\ldots k;
\ee
 $C_i$, $i=1,\ldots, k$ - $N\times N$ matrices with eigenvalues $e^{y^{(i)}_j}$, $j=1,\ldots,N$. Then
\[
 F^{\textsc{e},k;p}\left({a_1,\ldots,a_p}{n_1,\ldots, n_p}\,\bigg|\,\mathbf p^{(1)},\ldots,\mathbf p^{(k)}\right)=
\]
\[
 =\frac{1}{N!}\sum_{h_1,\ldots,h_N}\rule{-10pt}{10pt}'\;\;
 \prod_{a<b}^N \frac{|h_a-h_b|^{\beta}}{\prod_{i=1}^k (e^{y_a^{(i)}}-e^{y_b^{(i)}})}
 \prod_{m=1}^N(h_m!)^{-\beta} e^{-V_{h_m}}e^{y_m^{(1)}h_m}
 \prod_{i=2}^k \Theta_{\{h\}} 
 \det [ e^{y_j^{(i)}h_r} ]_{j,r\le N}.
\]

 \er

\br (A)
Let us note that the specification of the generating series (\ref{gen-hyp-funct-Pochhammer})
\[
F^{2,k;p}\left({a_1,\dots,a_p \atop 1,\dots, 1}{|\atop |}\bpow^{(1)},\dots,\bpow^{(k)}   \right)
\]
was considered in \cite{AMMN-2014} and denoted by $Z_{k,p}(a_1,\dots,a_p|\bpow^{(1)},\dots,\bpow^{(k)}) $ there.
In case $k=2$ and all $a_\alpha$ are natural numbers this is an example of the hypergeometric tau function, and in 
\cite{AMMN-2014} it was presented in form of the partition function of a certain chain-matrix model.

(B) The specification
\be\label{Gross-Richards}
_{p}F_q\left({a_1,\dots,a_p \atop b_1,\dots, b_q }{|\atop |}\bpow^{(1)},\bpow^{(2)}   \right ):=
F^{2,2;p+q}\left({a_1,\dots,a_p,b_1,\dots, b_q \atop 1,\dots, 1,-1,\dots, -1}{|\atop |}\bpow^{(1)},\bpow^{(2)}   \right)
\ee
may be identified with the so-called hypergeometric function of matrix argument (the case $\mathbb{C}$) introduced in \cite{Richards}
(the identification (\ref{Gross-Richards}) is correct only for all $a_i$ different otherwise the formula is more spacious).
It is an example of the TL tau functions studied in \cite{OS-2000}. In the context of Hurwitz numbers it was considered in
\cite{HO-2014}. The case $q=0$ describing the generation of  Goulden-Jackson Hurwitz numbers 
((\ref{GJ-Hurwitz-number'}) where $\textsc{e}=2$ and $k=1$ ) and it's analogue related to the Toda lattice hierarchy 
((\ref{GJ-Hurwitz-number'}) where $\textsc{e}=2$ and $k=2$ ) is of the special
importance because jointly with it's projective analogue 
\be\label{projective-Gross-Richards}
 _{p}{\tilde F}_q\left({a_1,\dots,a_p \atop b_1,\dots, b_q }{|\atop |}\bpow   \right ):=
F^{1,1;p+q}\left({a_1,\dots,a_p,b_1,\dots, b_q \atop 1,\dots, 1,-1,\dots, -1}{|\atop |}\bpow  \right)
\ee
written down in \cite{NO-2014}, \cite{NO-LMP}
 it contains all information about Hurwitz numbers which may be obtained with the help of intgerable systems. Matrix models
 giving rise to the perturbation series (\ref{Gross-Richards})
were considered in \cite{AMMN-2014}, in \cite{ChekhovAmbjorn} and will be considered below (we note, that all mentioned matrix models 
are different). Matrix models related to series
(\ref{projective-Gross-Richards}) are presented in \cite{NO-2014}, \cite{NO-LMP} and 
will be also presented in this paper
below.
Let us note that ${\rm Deg}\,_{p}F_q =2$ and ${\rm Deg}\,_p{\tilde F}_q = 1$.

(C) The specifiacation of the previous example where $p=0,\, q=1$ coincides with HCIZ integral and generates the so-called monotone 
Hurwitz numbers as it was considered in \cite{Goulden-Paquet-Novak}.

(D) The specification of (B) where $p=1,\, q=0$ coincides with asymptotic perturbation series for the two-matrix model obtained in \cite{HO-2003}. The further
specification $\bpow^{(2)}=\bpow_\infty$ yields the generating series for the so-called Belyi curves \cite{Chekhov-2014}, \cite{Zog}, 
\cite{KZ}.

\er

\bl\label{specialization-of-p} Let $k>n$.
\be
F^{\textsc{e},k-n;n}\left({a_1,\dots,a_n \atop 1,\dots,1 }{|\atop |}\bpow^{(n+1)},\dots,\bpow^{(k)}   \right)=
F^{\textsc{e},k;0}\left(\bpow^{(1)}(a_1),\dots,\bpow^{(n)}(a_n),\bpow^{(n+1)}\dots,\bpow^{(k)}  \right)
\ee
\el

 From written above it follows 
 \bp
 \[
F^{\textsc{e},k;p}\left({a_1,\dots,a_p \atop n_1,\dots, n_p}{|\atop |}\bpow^{(1)},\dots,\bpow^{(k)}   \right)=
\]
\be
\sum_{d>0}
\sum_{\Delta^1,\dots,\Delta^k\atop |\Delta^1|=\cdots=|\Delta^k|=d} \sum_{k_1,\dots , k_p}\,
G^{\textsc{e},k;p}\left(N;d;\Delta^1,\dots,\Delta^k {|\atop |} {k_1,\dots,k_p \atop n_1,\dots, n_p}\right)\,
\prod_{j=1}^p a_j^{dn_j-k_j} \,\prod_{i=1}^k \bpow^{(i)}_{\Delta^i}
 \ee
\ep

All matrix integrals below yields the perturbation series which are examples of this hypergeometric function generating  Hurwitz numbers
 (\ref{Hurwitz-number}), (\ref{GJ-Hurwitz-number}) and (\ref{gen-Hurwitz-number}).

\section{Integrals of functions of products of complex matrices \label{Complex}}

Below for simplicity we consider only $N\times N$ square matrices, however many results may be generalized also
for rectangular matrices. 

\br\label{free-p}
Let us note that the integrals considered in the present paper contain the set of
parameters denoted by ${\bf p}=(p_1,p_2,\dots)$. This is an important point. There exists domains of  the values of these parameters 
where integrals which will be considered converge. To see it one can take
 $p_i=-\sum_i^L y_k^i,\,i=1,2,\dots$ where each $y_i$ is a free parameter, then for any $L$ the series turns to be polynomials because
 $s_\lambda(\bpow)=0,\,\lambda=(\lambda_1,\lambda_2,\dots)$ for $\lambda_1>L$ as it follows from (\ref{ell<N}) and
 from (\ref{p-to-p-in-Schur}) in the Appendix. On the other hand, as we shall see the sums over partitions
 will be restricted in the lengths as $\ell(\lambda)\le N$, where $N$ is the matrix size, see (\ref{ell<N}).
\er

In the integrals of this Section the integration measure is defined by
\be\label{measure-Z-product}
d\Omega =\prod_{\alpha=1}^n d\mu(Z_\alpha)
\ee

For relations of this Section we use the notations
where $\tZ$, $\tZ^*$ and $\tZ^{(t)}$ which are given by (\ref{Z}), (\ref{Z^*}) and (\ref{tildeZ^*})

\subsection{Even Euler characteristic of the base. Complex matrices  \label{Even-complex}}

In this subsection we obtain simplest examples of the generalized hypergeometric function (\ref{gen-hyp-funct}).

\paragraph{Euler characteristic}  $\textsc{e}=2$,   {\bf and $n+2$ branch points}.
Let us consider the integral of the product of two simplest tau functions:
\bp\label{prop1}
\be\label{integral-1}
\int\tau_1^{\rm TL}(\tZ,\bpow)\tau_1^{\rm TL}(\tZ^*,\bpow^*)d\Omega=
\sum_{\lambda\atop \ell(\lambda)\le N} s_\lambda(\bpow)s_\lambda(\bpow^*)
\prod_{\alpha=1}^n\frac{s_\lambda(C_\alpha)}{s_\lambda(\bpow_\infty)}
\ee
\be
=\sum_{d>0}\sum_{\Delta^1\atop |\Delta^1|=d}\cdots
\sum_{\Delta^{n+2}\atop |\Delta^{n+2}|=d} 
H^{2,n+2}_N(d;\Delta^1,\dots,\Delta^{n+2})\prod_{\alpha=1}^{n+2}\bpow^{(\alpha)}_{\Delta^\alpha}
\ee
 where $\bpow^{(\alpha)}=(p^{(\alpha)}_1,p^{(\alpha)}_2,\dots),\,\alpha=1,\dots,n+2$ and we put 
$\bpow^{(n+1)}=\bpow,\quad \bpow^{(n+2)}=\bpow^*$ and
$$p^{(\alpha)}_m=\tr (C_\alpha)^m,\quad 1\le \alpha \le n $$

\ep

The right hand side of (\ref{integral-1}) is $F^{2,n+2;0}$ denoted by $Z_{n+2,0}$ in \cite{AMMN-2014}.

The sketch of the proof: First we apply (\ref{sAZZ^+B'}), then (\ref{sAZBZ^+'}) is applied $n$ times,
each time one gets the factor $\frac{s_\lambda(C_i)}{s_\lambda(\bpow_\infty)},\, i=1,\dots,n$. 
The degree ${\rm Deg}$ of the integrand is 4 (see Remark \ref{Deg-tau}), therefore due to the Remarks \ref{Deg} 
the Euler characteristic $\textsc{e}$ is 2.

\begin{Corollary}
If $C_\alpha=\mathbb{I}_N\,,\alpha=1,\dots,n$ then we obtain
\be\label{(N)-lambda-n}
\int e^{V(\tZ,\bpow)+V(\tZ^\dag,\bpow^*)} d\Omega =\sum_{\lambda} \left((N)_\lambda\right)^n s_\lambda(\bpow)s_\lambda(\bpow^*)
\ee
\end{Corollary}
The right hand side of (\ref{(N)-lambda-n}) is an example of the TL hypergeometric tau function \cite{OS-2000}
(in our notations this is an example of $F^{2,2;n}$, see Lemma \ref{specialization-of-p}).
 Let us note that this tau function was derived from
different multimatrix model: a chain of Hermitian matrices in \cite{AMMN-2014}.

Next, we consider the integral of the single simplest tau function:
\bp\label{prop2} With the notations of Proposition \ref{prop1} we have
\be\label{integral-2}
\int\tau_1^{\rm TL}(\tZ\tZ^*,\bpow) d\Omega=
\sum_{\lambda\atop \ell(\lambda)\le N} s_\lambda(\bpow) s_\lambda(\bpow_\infty)
\prod_{\alpha=1}^n\frac{s_\lambda(C_\alpha)}{s_\lambda(\bpow_\infty)}
\ee
\be
=\sum_{d>0}\sum_{\Delta^1\atop |\Delta^1|=d}\cdots
\sum_{\Delta^{n+1}\atop |\Delta^{n+1}|=d} H^{2,n+1}_N(d;\Delta^1,\dots,\Delta^{n+1})
\prod_{\alpha=1}^{n+1}\bpow^{(\alpha)}_{\Delta^\alpha}
\ee
The sum in the right hand side of (\ref{integral-2}) is the TL tau function in case 
$C_\alpha=\mathbb{I}_N,\,\alpha=1,\dots,n-1$ where
TL higher times are $\bpow^{(n)}$ and $\bpow^{(n+1)}$.
\ep
The right hand side of (\ref{integral-2}) is $F^{2,n+1;0}$.

The sketch of the proof: We apply the relation (\ref{sAZBZ^+'})  $n$ times,
each time getting the factor $\frac{s_\lambda(C_i)}{s_\lambda(\bpow_\infty)},\, i=1,\dots,n$. 
The degree ${\rm Deg}$ of the integrand is 2 (see Remark \ref{Deg-tau}), therefore due to the Remark \ref{Deg} 
the Euler characteristic $\textsc{e}$ is 2.

\begin{Corollary}
If $C_\alpha=\mathbb{I}_N\,,\alpha=1,\dots,n$ then we obtain
\be\label{(N)-lambda-n'}
\int e^{V(\tZ\tZ^\dag,\bpow)} d\Omega =\sum_{\lambda} \left((N)_\lambda\right)^n s_\lambda(\bpow)s_\lambda(\bpow_\infty)
\ee
\end{Corollary}
The integral in the right hand side of (\ref{integral-2}) where $C_\alpha=\mathbb{I}_N\,,\,\,\alpha=1,\dots,n-1$ was considered in work 
Ambjorn and Chekhov
\cite{ChekhovAmbjorn} where
Feynman graph technique for this integral was developed and the relation of the Feynman series to Hurwitz numbers was 
established. In this case the integral is equal to $F^{2,2;n-1}$ (see Lemma \ref{specialization-of-p}).

Let us replace the simplest tau function $\tau_1$ by the hypergeometric tau function $\tau_r$ of (\ref{hyp-tau}):
\bp\label{tau-1,2-r}
We have
\be\label{integral-1-r}
\int\tau_r^{\rm TL}(\tZ,\bpow)\tau_1^{\rm TL}(\tZ^*,\bpow^*)d\Omega=
\sum_{\lambda\atop \ell(\lambda)\le N} r_\lambda(n)s_\lambda(\bpow)s_\lambda(\bpow^*)
\prod_{\alpha=1}^n\frac{s_\lambda(C_\alpha)}{s_\lambda(\bpow_\infty)}
\ee
\be\label{integral-2-r}
\int\tau_r^{\rm TL}(\tZ\tZ^*,\bpow) d\Omega=
\sum_{\lambda\atop \ell(\lambda)\le N} r_\lambda(n) s_\lambda(\bpow) s_\lambda(\bpow_\infty)
\prod_{\alpha=1}^n\frac{s_\lambda(C_\alpha)}{s_\lambda(\bpow_\infty)}
\ee
\ep
\br\label{1-to-r}
Results of \cite{NO-LMP} concerning content products $r_\lambda$ allow the interpret series  
(\ref{integral-1-r}) and (\ref{integral-2-r}) as generating series for various weighted sums
of Hurwitz numbers where the weights are defined by the choice of the function $r$.
Due to Remark \ref{Deg-tau} the replacement $\tau_1 \to \tau_r$ does not change the Euler characteristic
$\textsc{e}$ of the base surface (which is Riemann sphere in examples (\ref{integral-1-r}) and (\ref{integral-2-r})).
\er

Example.  Take 
\be\label{r-rational}
 r(x)=\frac{\prod_{i=1}^p(a_i+x)}{\prod_{i=1}^q (b_i+x) }
\ee

Then in notations of Proposition \ref{prop1} we get
\[
\int\tau_r^{\rm TL}(\tZ,\bpow)\tau_1^{\rm TL}(\tZ^*,\bpow^*)d\Omega  = 
F^{2,n+2;p+q}\left({a_1,\dots,a_p, b_1,\dots, b_q \atop 1,\dots, 1,-1,\dots, -1}{|\atop |}\bpow^{(1)},\dots,\bpow^{(n+2)}  \right),
\]
and in notations of Proposition \ref{prop2} we get
\[
 \int\tau_r^{\rm TL}(\tZ\tZ^*,\bpow) d\Omega = 
F^{2,n+1;p+q}\left({a_1,\dots,a_p, b_1,\dots, b_q \atop 1,\dots, 1,-1,\dots, -1}{|\atop |}\bpow^{(1)},\dots,\bpow^{(n+1)}  \right)
\]

Next we produce the replacement $\tZ^* \to \tZ^{(t)}$ where $t>0$ and get different series which generate Hurwitz numbers with
different Euler chatacteristics of the base surfaces:

\paragraph{Euler characteristic} $\textsc{e} < 2$.

In previous example we mainly use the relation (\ref{sAZBZ^+'}).
In case we change the order of matrices in the product and instead of (\ref{Z^*}) we use (\ref{tildeZ^*}) then
we mostly use the relation (\ref{sAZZ^+B'}). Due to Remarks \ref{Euler=Deg} and \ref{Deg} it yields generating series
for rather different Hurwitz numbers:

\bp
(A) Let $t=2k \ge 2$. Introduce 
\be\label{C-odd-even}
C_{\rm odd}= C_1C_3 \cdots C_{2k-1} ,
\qquad
C_{\rm even}=C_2C_4\cdots C_{2k}
\ee

Then
\be\label{integral-3}
\int\tau_1^{\rm TL}(\tZ,\bpow^{(1)})\tau_1^{\rm TL}(\tZ^{(t)},\bpow^{(2)})d\Omega=
\sum_{\lambda\atop \ell(\lambda)\le N} 
\left( \frac{{\rm dim}\lambda}{d!} \right)^{4-2k}
\frac{s_\lambda(\bpow^{(1)})s_\lambda(\bpow^{(2)})s_\lambda(C_{\rm odd})
s_\lambda(C_{\rm even})   }{\left(s_\lambda(\bpow_\infty)\right)^4} 
\prod_{\alpha=t+1}^{n} \frac{s_\lambda(C_\alpha)}{s_\lambda(\bpow_\infty)}
\ee
\be
=\sum_{d>0}\sum_{\Delta^1\atop |\Delta^1|=d}\cdots
\sum_{\Delta^{4+n-2k}\atop |\Delta^{4+n-2k}|=d} H^{4-2k,4+n-2k}_N(d;\Delta^1,\dots,\Delta^{4+n-2k})
\prod_{\alpha=1}^{4+n-2k}\bpow^{(\alpha)}_{\Delta^\alpha}
\ee
where 
$$p^{(3)}_m=\tr (C_{\rm odd})^m,\qquad p^{(4)}_m=\tr (C_{\rm even})^m,
\qquad p^{(\alpha)}_m=\tr C_{\alpha -4+2k}^m,\quad\alpha= 5,\dots, 4+n-2k$$

(In all formulas we mean that the product $ \prod_{\alpha = t + 1}^n $ becomes one for $ t = n $.)

(B) Let $t=2k-1$ and now we use
\be\label{C-odd-even'}
C_{\rm odd}= C_1C_3 \cdots C_{2k-1} ,
\qquad
C_{\rm even}=C_2C_4\cdots C_{2k-2}
\ee

 Then
\be\label{integral-4}
\int\tau_1^{\rm TL}(\tZ,\bpow^{(1)})\tau_1^{\rm TL}(\tZ^{(t)},\bpow^{(2)})d\Omega=
\sum_{\lambda\atop \ell(\lambda)\le N} 
\left( \frac{{\rm dim}\lambda}{d!} \right)^{4-2k} 
\frac{s_\lambda(\bpow^{(1)})s_\lambda(\bpow^{(2)})
s_\lambda( C_{\rm odd}C_{\rm even}  )   }{\left(s_\lambda(\bpow_\infty)\right)^3}
\prod_{\alpha=t+1}^{n} \frac{s_\lambda(C_\alpha)}{s_\lambda(\bpow_\infty)}
\ee
\be
=\sum_{d>0}\sum_{\Delta^1\atop |\Delta^1|=d}\cdots
\sum_{\Delta^{4+n-2k}\atop |\Delta^{4+n-2k}|=d} H^{4-2k,4+n-2k}_N(d;\Delta^1,\dots,\Delta^{4+n-2k})
\prod_{\alpha=1}^{4+n-2k}\bpow^{(\alpha)}_{\Delta^\alpha}
\ee
where
$$p^{(3)}_m=\tr ( C_{\rm odd}C_{\rm even})^m,\qquad 
 p^{(\alpha)}_m=\tr C_{\alpha -4+2k}^m,\quad\alpha= 4,\dots, 4+n-2k
$$

\ep

The right hand side of (\ref{integral-3}) and the right hand side of (\ref{integral-4}) ) are equal respectively to $F^{4-t,4+n-t ;0}$ 
and to $F^{4-t-1,3+n-t ;0}$.

The sketch of the proof: We apply the relation (\ref{sAZZ^+B'})  $k$ times. 
The degree ${\rm Deg}$ of the integrand is 4, therefore thanks to the Remark \ref{Deg} the Euler 
characteristic of the base $\textsc{e}$ is $4-2k$.

Also we get

\bp 

(A) Let $t=2k \ge 2$. Then
\be\label{integral-5}
\int\tau_1^{\rm TL}(\tZ\tZ^{(t)},\bpow) d\Omega=
\sum_{\lambda\atop \ell(\lambda)\le N} 
\left( \frac{{\rm dim}\lambda}{d!} \right)^{4-2k}
\frac{s_\lambda(\bpow) s_\lambda(C_{\rm odd})s_\lambda(C_{\rm even})   }{\left(s_\lambda(\bpow_\infty)\right)^3}
\prod_{\alpha=t+1}^{n} \frac{s_\lambda(C_\alpha)}{s_\lambda(\bpow_\infty)}
\ee
\be
=\sum_{d>0}\sum_{\Delta^1\atop |\Delta^1|=d}\cdots
\sum_{\Delta^{3+n-2k}\atop |\Delta^{3+n-2k}|=d} H^{4-2k,3+n-2k}_N(d;\Delta^1,\dots,\Delta^{3+n-2k})
\prod_{\alpha=1}^{3+n-2k}\bpow^{(\alpha)}_{\Delta^\alpha}
\ee
 where  $\bpow^{(1)}=\bpow$ and where $C_{\rm odd}$ and $C_{\rm even}$ are given by (\ref{C-odd-even}) and where
$$p^{(2)}_m=\tr (C_{\rm odd})^m,\qquad p^{(3)}_m=\tr (C_{\rm even})^m, \qquad p^{(\alpha)}_m=
\tr C_{\alpha -3+2k}^m,\quad \alpha= 4,\dots,3+n-2k $$

(B) Let $t=2k-1 \ge 1$ and now $C_{\rm odd}$ and $C_{\rm even}$ are given by (\ref{C-odd-even'}). Then
\be\label{integral-6}
\int\tau_1^{\rm TL}(\tZ\tZ^{(t)},\bpow) d\Omega=
\sum_{\lambda\atop \ell(\lambda)\le N} 
\left( \frac{{\rm dim}\lambda}{d!} \right)^{4-2k}
\frac{s_\lambda(\bpow) s_\lambda(C_{\rm odd} C_{\rm even} ) }{\left(s_\lambda(\bpow_\infty)\right)^2}
 \prod_{\alpha=t+1}^{n} \frac{s_\lambda(C_\alpha)}{s_\lambda(\bpow_\infty)}
\ee
\be
=\sum_{d>0}\sum_{\Delta^1\atop |\Delta^1|=d}\cdots
\sum_{\Delta^{3+n-2k}\atop |\Delta^{3+n-2k}|=d} H^{4-2k,3+n-2k}_N(d;\Delta^1,\dots,\Delta^{3+n-2k})
\prod_{\alpha=1}^{3+n-2k}\bpow^{(\alpha)}_{\Delta^\alpha}
\ee
 where $\bpow^{(1)}=\bpow$ and
$$p^{(2)}_m=\tr (C_{\rm odd}C_{\rm even})^m, \qquad p^{(\alpha)}_m=
\tr C_{\alpha -3+2k}^m,\quad \alpha= 3,\dots,3+n-2k $$

\ep

At last, we consider the integral which may be interesting in the context of quantum chaos \cite{Ak1}:
\bp
Let
\[
Z'=\left(Z_1C_1Z_1^\dag A_{12}  \right)\cdots \left(Z_nC_nZ_n^\dag A_{n1}  \right)
\]
Then
\be\label{integral-6-Chekhov-Strahov}
\int\tau_1^{\rm TL}(Z',\bpow) d\Omega=
\sum_{\lambda\atop \ell(\lambda)\le N} s_\lambda(\bpow) s_\lambda(A_{12}\cdots A_{n1})
\prod_{\alpha=1}^n\frac{s_\lambda(C_\alpha)}{s_\lambda(\bpow_\infty)}
\ee
\be
=\sum_{d>0}\sum_{\Delta^1\atop |\Delta^1|=d}\cdots
\sum_{\Delta^{n+2}\atop |\Delta^{n+2}|=d} H^{2,n+2}_N(d;\Delta^1,\dots,\Delta^{n+2})
\prod_{\alpha=1}^{n+2}\bpow^{(\alpha)}_{\Delta^\alpha}
\ee
where
$\bpow^{(\alpha)}=(p^{(\alpha)}_1,p^{(\alpha)}_2,\dots),\,\alpha=1,\dots,n+2$ and where we put 
$\bpow^{(n+1)}=\bpow,\quad p^{(n+2)}_m=\tr \left(A_{12}\cdots A_{n1}  \right)^m$ and where
$$p^{(\alpha)}_m=\tr (C_\alpha)^m,\quad 1\le \alpha \le n $$

\ep

Let us note that the right hand side in formula (\ref{integral-6-Chekhov-Strahov}) coincides with
 the generating series denoted  by $Z_{n+2,0}$  in \cite{AMMN-2014}.

\br\label{order-change-remark-Z}
If one introduces
\[
  \tZ^\sigma = Z^\dag_{\sigma(n)} Z^\dag_{\sigma(n-1)} \cdots Z^\dag_{\sigma(1)}
\]
where $ \sigma \in S_n $, then the integral of $ \tau_1 (\tZ \tZ^{\sigma}) $ is the generating function for Hurwitz numbers,
which will be written out in another paper. Note that such an integral can be associated with a chord diagram,  whoose $ 2n $ vertices
are numbered by the sequence of numbers $ 1,2, \dots, n, \sigma(n), \sigma(n-1), \dots, \sigma(1) $, and vertices with the same
numbers are connected by chords. The chords correspond to the pairing of the matrices 
$ (Z_\alpha, Z_\alpha^\dag), \, \alpha = 1, \dots, n $ by the rules,
given by the equations (\ref{sAZBZ^+'}) and (\ref{sAZZ^+B'}).
\er

\br
Let us note that
here and below throught the paper one can replace the simplest tau function $\tau_1$ by the hypergeometric
tau functions $\tau_r$ and, in such a way, get the generating function for more general weighted sums of Hurwitz numbers
where weights are defiened by the choice of the function $r$.
\er

\subsection{Odd Euler characteristic of the base. Complex matrices \label{Odd-complex}}

In this subsection we present generating functions Hurwitz numbers of Klein surfaces.

\paragraph{Euler characteristic}  $\textsc{e}=1$, {\bf and $n+1$ branch points}.
\bp\label{prop1-odd}
\be\label{integral-1-odd}
\int\tau_1^{\rm TL}(\tZ,\bpow)\tau_1^{\rm BKP}(\tZ^*)d\Omega=
\sum_{\lambda\atop \ell(\lambda)\le N} s_\lambda(\bpow)
\prod_{\alpha=1}^n\frac{s_\lambda(C_\alpha)}{s_\lambda(\bpow_\infty)}
\ee
\be
=\sum_{d>0}\sum_{\Delta^1\atop |\Delta^1|=d}\cdots
\sum_{\Delta^{n+1}\atop |\Delta^{n+1}|=d} 
H^{1,n+1}_N(d;\Delta^1,\dots,\Delta^{n+1})\prod_{\alpha=1}^{n+1}\bpow^{(\alpha)}_{\Delta^\alpha}
\ee
where $\tZ$ and $\tZ^*$ are given respectively by (\ref{Z}) and (\ref{Z^*}), where the measure
\be
d\Omega =\prod_{\alpha=1}^n d\mu(Z_\alpha)
\ee
and where $\bpow^{(\alpha)}=(p^{(\alpha)}_1,p^{(\alpha)}_2,\dots),\,\alpha=1,\dots,n+2$ and we put 
$\bpow^{(n+1)}=\bpow $ and
$$p^{(\alpha)}_m=\tr (C_\alpha)^m,\quad 1\le \alpha \le n $$

\ep

\begin{Corollary}
If $C_\alpha=\mathbb{I}_N\,,\alpha=1,\dots,n$ then we obtain
\be
\int e^{V(\tZ,\bpow)}\tau_1^{\rm BKP}(\tZ^*) d\Omega =\sum_{\lambda} \left((N)_\lambda\right)^n s_\lambda(\bpow)
\ee
The right hand side is an example of the BKP hypergeometric tau function \cite{OST-I}. 

\end{Corollary}

  \br\label{prop2-odd} With the notations of Proposition \ref{prop1-odd} we formaly have
  \be\label{integral-2-odd}
  \int\tau_1^{\rm BKP}(\tZ\tZ^*) d\Omega=
  \sum_{\lambda\atop \ell(\lambda)\le N} s_\lambda(\bpow_\infty)
 \prod_{\alpha=1}^n\frac{s_\lambda(C_\alpha)}{s_\lambda(\bpow_\infty)}
  \ee
  \be
  =\sum_{d>0}\sum_{\Delta^1\atop |\Delta^1|=d}\cdots
  \sum_{\Delta^{n}\atop |\Delta^{n}|=d} H^{1,n}_N(d;\Delta^1,\dots,\Delta^{n})
  \prod_{\alpha=1}^{n}\bpow^{(\alpha)}_{\Delta^\alpha}
  \ee
  which should be interpreted as a term by term equality if one developes the integrand in the Schur functions, see (\ref{tau_1^B}). 
  Here, we have no the set of free parameters $\bpow$ as we have it in other integrals, see Remark \ref{free-p},
  and we can not terminate the series 
  by a special choice of these parameters.
  For
  $n=1$ the right hand side is the know expression for the sum of the Schur functions, see \cite{Mac}. For $n>1$
  the sum in the right hand side of (\ref{integral-2-odd}) is the BKP tau function in case $C_\alpha=\mathbb{I}_N,\,\alpha=1,\dots,n-1$ where
  BKP higher times are $\bpow^{(n)}$.
 \er

Next we produce the replacement $\tZ^* \to \tZ^{(t)}$ similar to the previous case where we considered the tau functions of the
TL hierarchy,
and get different series which generate Hurwitz numbers with
different Euler chatacteristics of the base surfaces:

\paragraph{Euler characteristic} $\textsc{e} < 1$.

\bp
(A) Let $t=2k \ge 2$. 
Then
\be
\int\tau_1^{\rm TL}(\tZ,\bpow^{(1)})\tau_1^{\rm BKP}(\tZ^{(t)})d\Omega=
\sum_{\lambda\atop \ell(\lambda)\le N} 
\left( \frac{{\rm dim}\lambda}{d!} \right)^{3-2k}
\frac{s_\lambda(\bpow^{(1)}) s_\lambda(C_{\rm odd})s_\lambda(C_{\rm even})   }{\left(s_\lambda(\bpow_\infty)\right)^3}
\prod_{\alpha=t+1}^{n} \frac{s_\lambda(C_\alpha)}{s_\lambda(\bpow_\infty)}
\ee
\be
=\sum_{d>0}\sum_{\Delta^1\atop |\Delta^1|=d}  \cdots
\sum_{\Delta^{3+n-2k}\atop |\Delta^{3+n-2k}|=d} H^{3-2k,3+n-2k}_N(d;\Delta^1,\dots,\Delta^{3+n-2k})
\prod_{\alpha=1}^{3+n-2k}\bpow^{(\alpha)}_{\Delta^\alpha}
\ee
where   
$$p^{(2)}_m=\tr (C_{\rm odd})^m,\qquad p^{(3)}_m=\tr (C_{\rm even})^m, \qquad
p^{(\alpha)}_m=\tr C_{\alpha -3+2k}^m,\quad\alpha= 4,\dots, 3+n-2k
$$
Here $C_{\rm odd}$ and $C_{\rm even}$ are given by (\ref{C-odd-even}).

(B) Let $n=2k-1$. Now $C_{\rm odd}$ and $C_{\rm even}$ are given by (\ref{C-odd-even'}). Then
\be
\int\tau_1^{\rm TL}(\tZ,\bpow^{(1)})\tau_1^{\rm BKP}(\tZ^{(t)})d\Omega=
\sum_{\lambda\atop \ell(\lambda)\le N} 
\left( \frac{{\rm dim}\lambda}{d!} \right)^{3-2k}
\frac{s_\lambda(\bpow^{(1)}) s_\lambda(C_{\rm odd}C_{\rm even})   }{\left(s_\lambda(\bpow_\infty)\right)^2}
\prod_{\alpha=t+1}^{n} \frac{s_\lambda(C_\alpha)}{s_\lambda(\bpow_\infty)}
\ee
\be
=\sum_{d>0}\sum_{\Delta^1\atop |\Delta^1|=d}\cdots
\sum_{\Delta^{3+n-2k}\atop |\Delta^{3+n-2k}|=d} H^{3-2k,3+n-2k}_N(d;\Delta^1,\cdots,
\Delta^{3+n-2k})\prod_{\alpha=1}^{3+n-2k}\bpow^{(\alpha)}_{\Delta^\alpha}
\ee
where 
$$p^{(2)}_m=\tr (C_{\rm odd}C_{\rm even})^m , \qquad
p^{(\alpha)}_m=\tr C_{\alpha -3+2k}^m,\quad\alpha= 3,\dots, 3+n-2k$$

\ep
The right hand sides of (\ref{integral-3}) and of (\ref{integral-4}) are equal respectively to $F^{3-t,3+n-t ;0}$ and to $F^{2-t,2+n-t ;0}$.

The sketch of the proof: We apply the relation (\ref{sAZZ^+B'})  $k$ times. 
The degree ${\rm Deg}$ of the integrand is 3 (see Remark \ref{Deg-tau}), therefore thanks to the Remark \ref{Deg} the Euler 
chacteristic $\textsc{e}$ is $3-2k$.

\br
The replacememnt
\be
d\Omega\,\to\,d\Omega(a_1,\dots,a_k) =\prod_{i=1}^k \det(Z_\alpha Z^\dag_\alpha)^{a_i}  d\mu(Z_i)
\ee
result to the multiplication factor inside each term inside the sum over partitions $\lambda$
 by $\prod_{\alpha=1}^n (a_i+N)_\lambda$. See Section \ref{tau-functions-2KP-section} for more details.
 Compare to \cite{Ak1} and also to a different matrix model - a chain matrix model introduces in \cite{AMMN-2014}.
\er

\subsection{Representation of (\ref{(N)-lambda-n}) in form of the fermionic vacuum expectation value}

Each hepergeometric tau function in both either the KP or BKP cases may be presented in form
of a fermionic expectation value. Let us show the simplest case (\ref{(N)-lambda-n}) as an example.

Two-component Fermi fields are denoted by $\psi^{(i)}$ and $\psi^{\dag(i)}$, where $i=1,2$. The right vacuum is
$|0,0\rangle$, the left vacuum, $\langle N,-N|$, has different levels of Dirac seas which is $N$ for the first component and
 $-N$ for the second, details see in Section  \ref{Product-of-matrices-and-2KP} in Appendix.

\bp
 We have
\[
\sum_\lambda \left( (N)_\lambda\right)^n s_\lambda(\bpow)s_\lambda(\bpow^*) = 
\]
\be\label{tau-Fermi}
\int \prod_{N\ge i>j\ge 1}|z_i-z_j|^2 
\prod_{i=1}^N e^{V(z_i,t)+V({\bar z}_i, p^*)} w(n,|z_i|)d^2 z_i
= \langle N,-N| \Gamma^{(1)}(p)\Gamma^{(2)}(p^*) g |0,0\rangle
\ee
where
\be
\Gamma^{(\alpha)}(p) = \exp \sum_{m>0} p_m J_m^{(\alpha)},\quad    
J_m^{(\alpha)}=\sum_{k\in\mathbb{Z}} \psi^{(\alpha)}_k \psi^{\dag(\alpha)}_{k+m}
\ee
and
\be\label{g}
g=\exp \int \psi^{(1)}(z)\psi^{\dag(2)}({\bar z}) w(n,|z|) d^2z
\ee
where 
\be\label{w}
w(n,z) = \int_{}^{} \left( \Gamma(s)\right)^n |z|^{-2s} ds
\ee
\ep

Here $\Gamma(s)$ is the Gamma function (don't confuse with $\Gamma^{(\alpha)}$ above).

For the proof we use 
\be\label{r-via-moments}
\int |z|^{2k} w(n,|z|)d^2 z =r(1)\cdots r(k)
\ee
We need to verify that  if for $w(n,|z|)$ we choose (\ref{w}) we get $r(x)=x^n$.
Indeed
\be\label{w-property}
\int_{\gamma_0} \left( \Gamma(s)\right)^n |z|^{2k-2s} d^2z ds =
\int_{\gamma_k} \left( \Gamma(s+k)\right)^n |z|^{-2s} d^2z ds
=1^n2^n\cdots k^n\int_{\gamma_0} \left( \Gamma(s)\right)^n |z|^{-2s} d^2z ds
\ee
where the countor $\gamma_0$ in the complex $s$-plane is the line 
$\left(-i\infty + \frac 12,+i\infty +\frac 12\right)$ 
while
$\gamma_k = \left(-i\infty + \frac 12 -k,+i\infty +\frac 12 -k\right)$. We use
$\Gamma(s)\Gamma(1-s)=\frac{\pi}{\sin \pi s}$.

More information may be found in the mentioned Section \ref{Product-of-matrices-and-2KP} of the Appendix.

\section{Integrals of products of unitary matrices\label{Unitary}}

Consider the following products
\bea\label{U}
\tU &:=&(U_1 C_1) \cdots (U_n C_n)\\ 
\label{U^*}
\tU^* &:=& U_n^\dag\cdots U_1^\dag\\
\label{tildeU^*}
\tU^{(t)}&:=& U^\dag_n  \cdots U^\dag_{t+1} U^\dag_1\cdots U^\dag_t,\qquad t  \le n 
\eea
where $U_\alpha^\dag \in \mathbb{U}(N)$. (The matrix $\tU^{(t)}$ may be obtained 
by the action of the product of $[\frac t2]$ transpositions on $\tU^*$.)

We use the same tau functions $\tau_1(X,\bpow)$ and $\tau_1^{\rm BKP}(X)$ introduced by relations (\ref{tau_1})-(\ref{tau_1^B}).
In this paper $d_*U$ denotes the Haar measure on the unitary group,  for explicit formulae see (\ref{Haar-unitary}) in Appendix. 
We also recall that $s_\lambda(\mathbb{I}_N)=(N)_\lambda s_\lambda(\bpow_\infty)=\frac{{\rm dim}\lambda}{|\lambda|!}(N)_\lambda$.

Almost all results related to unitary matrices may be obtained from results related to complex matrices obtained in the previous section
by the replacement $s_\lambda(\bpow_\infty)\to s_\lambda(\mathbb{I}_N)$ which follows from the replacement
of relations (\ref{sAZBZ^+'})-(\ref{sAZZ^+B'}) by relations (\ref{sAUBU^+1})-(\ref{sAUU^+B'}). 

For simplicification of formulae we will consider only two special cases: $t=0$ and $t=n$.

\subsection{Even Euler characteristic of the base. Unitary matrices \label{Even-unitary} }

\paragraph{Euler characteristic}  $\textsc{e}=2$.
\bp\label{prop1-U}
\be\label{integral-1-U}
\int\tau_1^{\rm TL}(\tU,\bpow)\tau_1^{\rm TL}(\tU^*,\bpow^*)\prod_{\alpha=1}^n d_*U_\alpha=
\sum_{\lambda\atop \ell(\lambda)\le N} s_\lambda(\bpow)s_\lambda(\bpow^*)
\prod_{\alpha=1}^n\frac{s_\lambda(C_\alpha)}{s_\lambda(\mathbb{I}_N)}
\ee
\be
=F^{2,n+2;1}\left({N \atop -n}{|\atop |}\bpow^{(1)},\dots,\bpow^{(n+2)}  \right)
\ee
where $\tU$ and $\tU^*$ are given by (\ref{U}) and (\ref{U^*}), where
$ d_*U_\alpha $ is the Haar measure on $\mathbb{U}(N)$,
and where $\bpow^{(\alpha)}=(p^{(\alpha)}_1,p^{(\alpha)}_2,\dots),\,\alpha=1,\dots,n+2$ and we put 
$\bpow^{(n+1)}=\bpow,\quad \bpow^{(n+2)}=\bpow^*$ and
$$p^{(\alpha)}_m=\tr (C_\alpha)^m,\quad 1\le \alpha \le n $$

\ep

\begin{Corollary}
If $C_\alpha=\mathbb{I}_N\,,\alpha=1,\dots,n$ then we obtain
\[
\int e^{V(\tU,\bpow)+V(\tU^\dag,\bpow^*)} \prod_{\alpha=1}^n d_*U_\alpha =\sum_{\lambda}  s_\lambda(\bpow)s_\lambda(\bpow^*)
\]
which does not depend on $n$.

\end{Corollary}

Next integral:
\bp\label{prop2-U} With the notations of Proposition \ref{prop1-U} we have
\be\label{integral-2-U}
\int\tau_1^{\rm TL}(\tU\tU^*,\bpow) \prod_{\alpha=1}^n d_*U_\alpha=
\sum_{\lambda\atop \ell(\lambda)\le N} s_\lambda(\bpow) s_\lambda(\mathbb{I}_N)
\prod_{\alpha=1}^n\frac{s_\lambda(C_\alpha)}{s_\lambda(\mathbb{I}_N)}
\ee
\be
= F^{2,n+1;1}\left({N \atop 1-n}{|\atop |}\bpow^{(1)},\dots,\bpow^{(n+1)}  \right)
\ee
The sum in the right hand side of (\ref{integral-2-U}) is the TL tau function in case $C_\alpha=\mathbb{I}_N,\,\alpha=1,\dots,n-1$ where
TL higher times are $\bpow^{(n)}$ and $\bpow^{(n+1)}$.

\ep

Next we produce the replacement $\tU^* \to \tU^{(n)}$ and get different series which generate Hurwitz numbers with
different Euler chatacteristics of the base surfaces:

\paragraph{Euler characteristic} $\textsc{e}=4-2k$. The replacement $U^* \to {\tilde U}^*$ gives the results similar to
the case of complex matrices in Section \ref{Complex}:

\bp
Let $\tU$ and $\tU^{(n)}$ are given respectively by (\ref{U}) and (\ref{tildeU^*}).

(A) If $n=2k \ge 2$. Then
\be
\int\tau_1^{\rm TL}(\tU,\bpow^{(1)})\tau_1^{\rm TL}(\tU^{(n)},\bpow^{(2)})\prod_{\alpha=1}^n d_*U_\alpha=
\sum_{\lambda\atop \ell(\lambda)\le N} 
s_\lambda(\bpow^{(1)})s_\lambda(\bpow^{(2)})s_\lambda(C_1C_3\cdots C_{2k-1})s_\lambda(C_2C_4\cdots C_{2k}) 
\left( s_\lambda(\mathbb{I}_N) \right)^{-2k}
\ee
\be
 = F^{4-2k,4;1}\left({N \atop -2k}{|\atop |}\bpow^{(1)},\dots,\bpow^{(4)}  \right)
\ee
 where   
$$p^{(3)}_m=\tr (C_1C_3\cdots C_{2k-1})^m,\qquad p^{(4)}_m=\tr (C_2C_4\cdots C_{2k})^m$$

(B) If $n=2k-1$. Then
\be
\int\tau_1^{\rm TL}(\tU,\bpow^{(1)})\tau_1^{\rm TL}(\tU^{(n)},\bpow^{(2)})\prod_{\alpha=1}^n d_*U_\alpha=
\sum_{\lambda\atop \ell(\lambda)\le N} 
s_\lambda(\bpow^{(1)})s_\lambda(\bpow^{(2)})s_\lambda(C_1C_3\cdots C_{2k-1}C_2C_4\dots C_{2k-2})
\left(s_\lambda(\mathbb{I}_N)\right)^{1-2k}
\ee
\be
=F^{4-2k,3;1}\left({N \atop 1-2k}{|\atop |}\bpow^{(1)}, \bpow^{(2)}  ,\bpow^{(3)}  \right)
\ee
where 
$$p^{(3)}_m=\tr (C_1C_3\cdots C_{2k-1}C_2C_4\dots C_{2k-2})^m $$

\ep

We also get

\bp Let $\tU$ and $\tU^{(n)}$ as previousely are given by respectively (\ref{U}) and (\ref{tildeU^*}).

(A) Let $n=2k \ge 2$. Then
\be
\int\tau_1^{\rm TL}(\tU\tU^{(n)},\bpow) \prod_{\alpha=1}^n d_*U_\alpha=
\sum_{\lambda\atop \ell(\lambda)\le N} 
s_\lambda(\bpow) s_\lambda(C_1C_3\cdots C_{2k-1})s_\lambda(C_2C_4\cdots C_{2k}) \left(s_\lambda(\mathbb{I}_N) \right)^{1-2k}
\ee
\be
= F^{4-2k,3;1}\left({N \atop 1-2k}{|\atop |}\bpow^{(1)}, \bpow^{(2)}  ,\bpow^{(3)}  \right)
\ee
 where  $\bpow^{(1)}=\bpow$ and 
$$p^{(2)}_m=\tr (C_1C_3\cdots C_{2k-1})^m,\qquad p^{(3)}_m=\tr (C_2C_4\cdots C_{2k})^m$$

(B) Let $n=2k-1 \ge 1$. Then
\be
\int\tau_1^{\rm TL}(\tU\tU^{(n)},\bpow) \prod_{\alpha=1}^n d_*U_\alpha=
\sum_{\lambda\atop \ell(\lambda)\le N} 
s_\lambda(\bpow) s_\lambda(C_1C_3\cdots C_{2k-1}C_2C_4\dots C_{2k-2}) \left(s_\lambda(\mathbb{I}_N) \right)^{2-2k}
\ee
\be
=F^{4-2k,2;1}\left({N \atop 2-2k}{|\atop |}\bpow^{(1)}, \bpow^{(2)}  \right)
\ee
 where $\bpow^{(1)}=\bpow$ and
$$p^{(2)}_m=\tr (C_1C_3\cdots C_{2k-1}C_2C_4\dots C_{2k-2})^m $$

\ep

\br\label{order-change-remark-U-even}
Note that if we replace $U^*$ of the formula (\ref{U^*}) by the product
$U_{\sigma(n)}^\dag U_{\sigma(n-1)}^\dag \cdots  U_{\sigma(1)}^\dag$, we obtain  different Euler characteristic $\textsc{e}$,
which is determined by the choice of the permutation $\sigma\in S_n $. By replacing $U^{(n)}$ with $U^{(t)}$, we obtain  
$\textsc{e}=2-2[\frac 12 t]$.
\er

\subsection{Odd Euler characteristic of the base. Unitary matrices \label{Odd-unitary} }

In this subsection we present generating functions Hurwitz numbers of Klein surfaces.

\paragraph{Euler characteristic}  $\textsc{e}=1$, {\bf and $n+1$ branch points}.
\bp\label{prop1-odd-U}
\be\label{integral-1-odd-U}
\int\tau_1^{\rm TL}(\tU,\bpow)\tau_1^{\rm BKP}(\tU^*)\prod_{\alpha=1}^n d_*U_\alpha=
\sum_{\lambda\atop \ell(\lambda)\le N} s_\lambda(\bpow)
\prod_{\alpha=1}^n\frac{s_\lambda(C_\alpha)}{s_\lambda(\mathbb{I}_N)}
\ee
\be
=F^{1,n+1;1}\left({N \atop -n}{|\atop |}\bpow^{(1)},\dots,\bpow^{(n+1)}  \right)
\ee
where $\tU$ and $\tU^*$ are given respectively by (\ref{U}) and (\ref{U^*}), 
and where $\bpow^{(\alpha)}=(p^{(\alpha)}_1,p^{(\alpha)}_2,\dots),\,\alpha=1,\dots,n+1$. Here we put 
$\bpow^{(n+1)}=\bpow $ and
$$p^{(\alpha)}_m=\tr (C_\alpha)^m,\quad 1\le \alpha \le n $$

\ep

\begin{Corollary}
If $C_\alpha=\mathbb{I}_N\,,\alpha=1,\dots,n$ then we obtain
\be
\int e^{V(\tU,\bpow)}\tau_1^{\rm BKP}(\tU^*) \prod_{\alpha=1}^n d_*U_\alpha =\sum_{\lambda} \left((N)_\lambda\right)^n s_\lambda(\bpow)
\ee
which is an example of the BKP hypergeometric tau function \cite{OST-I}. 

\end{Corollary}

Next
\bp\label{prop2-odd-U} With the notations of Proposition \ref{prop1-odd-U} we have
\be\label{integral-2-odd-U}
\int\tau_1^{\rm BKP}(\tU\tU^*) \prod_{\alpha=1}^n d_*U_\alpha=
\sum_{\lambda\atop \ell(\lambda)\le N} s_\lambda(\mathbb{I}_N)
\prod_{\alpha=1}^n\frac{s_\lambda(C_\alpha)}{s_\lambda(\mathbb{I}_N)}
\ee
\be
= F^{1,n;1}\left({N \atop 1-n}{|\atop |}\bpow^{(1)},\dots,\bpow^{(n)}  \right)
\ee
The sum in the right hand side of (\ref{integral-2-odd-U}) is the BKP tau function in case $C_\alpha=\mathbb{I}_N,\,\alpha=1,\dots,n-1$ 
where the set $\bpow^{(n)}$ plays the role of higher times of the hierarchy.
\ep

Next we produce the replacement $\tU^* \to \tU^{(n)}$ and get series which generate Hurwitz numbers with
different Euler chatacteristics of the base surfaces:

\paragraph{Euler characteristic} $\textsc{e}=3-2k$, {\bf three and  two branch points}.

\bp
(A) Let $n=2k \ge 2$. Then
\be
\int\tau_1^{\rm TL}(\tU,\bpow^{(1)})\tau_1^{\rm BKP}(\tU^{(n)})\prod_{\alpha=1}^n d_*U_\alpha=
\sum_{\lambda\atop \ell(\lambda)\le N} 
\frac{s_\lambda(\bpow^{(1)}) s_\lambda(C_1C_3\cdots C_{2k-1})s_\lambda(C_2C_4\cdots C_{2k})   }{\left(s_\lambda(\mathbb{I}_N)\right)^3}
\left( s_\lambda(\mathbb{I}_N \right)^{3-2k}
\ee
\be
=F^{3-2k,3;1}\left({N \atop -2k}{|\atop |}\bpow^{(1)},\bpow^{(2)},\bpow^{(3)}  \right)
\ee
where $\tU$ and $\tU^{(n)}$ are given by (\ref{U}) and (\ref{tildeU^*}), 
and where   
$$p^{(2)}_m=\tr (C_1C_3\cdots C_{2k-1})^m,\qquad p^{(3)}_m=\tr (C_2C_4\cdots C_{2k})^m$$

(B) Let $n=2k-1$. Then
\be
\int\tau_1^{\rm TL}(\tU,\bpow^{(1)})\tau_1^{\rm BKP}(\tU^{(n)})\prod_{\alpha=1}^n d_*U_\alpha=
\sum_{\lambda\atop \ell(\lambda)\le N} 
\frac{s_\lambda(\bpow^{(1)}) s_\lambda(C_1C_3\cdots C_{2k-1}C_2C_4\dots C_{2k-2} )   }{\left(s_\lambda(\mathbb{I}_N)\right)^2}
\left( s_\lambda(\mathbb{I}_N \right)^{3-2k}
\ee
\be
=F^{3-2k,2;1}\left({N \atop 1-2k}{|\atop |}\bpow^{(1)}, \bpow^{(2)}   \right)
\ee
where $\tU$ and $\tU^{(n)}$ are given by (\ref{U}) and (\ref{tildeU^*}), 
and where 
$p^{(2)}_m=\tr (C_1C_3\cdots C_{2k-1}C_2C_4\dots C_{2k-2} )^m $.

\ep

Also

\bp Let $\tU$ and $\tU^{(n)}$ are given by respectively (\ref{U}) and (\ref{tildeU^*}).

(A) Let $n=2k \ge 2$. Then
\be\label{integral-5-odd-U}
\int\tau_1^{\rm BKP}(\tU\tU^{(n)}) \prod_{\alpha=1}^n d_*U_\alpha=
\sum_{\lambda\atop \ell(\lambda)\le N} 
\frac{ s_\lambda(C_1C_3\cdots C_{2k-1})s_\lambda(C_2C_4\cdots C_{2k})   }{\left(s_\lambda(\mathbb{I}_N)\right)^2}
\left( s_\lambda(\mathbb{I}_N \right)^{3-2k}
\ee
\be
=F^{3-2k,2;1}\left({N \atop 1-2k}{|\atop |}\bpow^{(1)}, \bpow^{(2)}  \right)
\ee
 where   
$$p^{(1)}_m=\tr (C_1C_3\cdots C_{2k-1})^m,\qquad p^{(2)}_m=\tr (C_2C_4\cdots C_{2k})^m$$

(B) Let $n=2k-1 \ge 1$. Then
\be\label{integral-6-odd-U}
\int\tau_1^{\rm BKP}(\tU\tU^{(n)}) \prod_{\alpha=1}^n d_*U_\alpha=
\sum_{\lambda\atop \ell(\lambda)\le N} 
\frac{s_\lambda(C_1C_3\cdots C_{2k-1}C_2C_4\dots C_{2k-2})   }{s_\lambda(\mathbb{I}_N)}
\left( s_\lambda(\mathbb{I}_N \right)^{3-2k}
\ee
\be
=F^{3-2k,1;1}\left({N \atop 2-2k}{|\atop |}\bpow \right)
\ee
 where 
$p_m=\tr (C_1C_3\cdots C_{2k-1}C_2C_4\dots C_{2k-2})^m $.

\ep
The right hand side of (\ref{integral-5-odd-U}) (and of (\ref{integral-6-odd-U}) ) is equal 
to $F^{3-2k,2 ;0}$ (respectively to $F^{3-2k,1 ;0}$).

\br\label{order-change-remark-U-odd}
If we replace $U^*$ of the formula (\ref{U^*}) by the product
$U_{\sigma(n)}^\dag U_{\sigma(n-1)}^\dag \cdots  U_{\sigma(1)}^\dag$, we obtain  different Euler characteristic $\textsc{e}$,
which is determined by the choice of the permutation $\sigma\in S_n $. By replacing $U^{(n)}$ with $U^{(t)}$, we obtain  
$\textsc{e}=1-2[\frac 12 t]$.
\er

\section{Discussion}

In this paper we considered matrix models of a certain type: these are integrals of hypergeometric tau functions, the 
product of matrices being the argument of the tau functions. We mainly considered the simplest ('vacuum') tau functions,
however it was shown what to do in the case of any tau functions of this type in Remark \ref{1-to-r}, see also examples given by
(\ref{r-rational}) of Proposition \ref{tau-1,2-r}. The tau functions of two hierarchies were considered: of the TL hierarchy (in particular, of KP) 
and of the BKP hierarchy of Kac-van-de Leur. We showed that such integrals generate Hurwitz numbers for covering problems of Riemann 
and of Klein surfaces with various Euler characteristics. The Euler characteristics are defined by the order of the matrices in the product. 
The integrals themselves typically are not tau functions themselves (and obviously aren't at not positive Euler characteristics $\textsc{e}$
of the base), though one can hope that the integrals inherit certain symmetries of tau functions. We used only complex
and unitary matrices, but it would be interesting to find other examples.

\section*{Acknowledgements}

The work has been funded by the RAS Program ``Fundemental problems of nonlinear mechanics'' and by the Russian Academic Excellence Project '5-100'.
I thank A. Odziyevich and university of Bialystok for warm hospitality which allowed
it is accurate to write down this work. I am grateful to S. Natanzon, A. Odziyevich, J. Harnad, A. Mironov (ITEP) and
to van de Ler for various remarks concerning the questions connected with this work. Special gratitude
to E. Strakhov for the fact that he drew my attention to the works on quantum chaos devoted to the products 
of random matrices and for fruitful discussions.

\appendix

\section{Partitions and Schur functions \label{Partitions-and-Schur-functions}}

Let us recall that the characters of the unitary group $\mathbb{U}(N)$ are labeled by partitions
and coincide with the so-called Schur functions \cite{Mac}. 
A partition 
$\lambda=(\lambda_1,\dots,\lambda_n)$ is a set of nonnegative integers $\lambda_i$ which are called
parts of $\lambda$ and which are ordered as $\lambda_i \ge \lambda_{i+1}$. 
The number of non-vanishing parts of $\lambda$ is called the length of the partition $\lambda$, and will be denoted by
 $\ell(\lambda)$. The number $|\lambda|=\sum_i \lambda_i$ is called the weight of $\lambda$. The set of all
 partitions will be denoted by $\mathbb{P}$.

The Schur function labelled by $\lambda$ may be defined as  the following function in variables
$x=(x_1,\dots,x_N)$ :
\be\label{Schur-x}
 s_\lambda(x)=\frac{\det \left[x_j^{\lambda_i-i+N}\right]_{i,j}}{\det \left[x_j^{-i+N}\right]_{i,j}}
 \ee
 in case $\ell(\lambda)\le N$ and vanishes otherwise. One can see that $s_\lambda(x)$ is a symmetric homogeneous 
 polynomial of degree $|\lambda|$ in the variables $x_1,\dots,x_N$, and $\deg x_i=1,\,i=1,\dots,N$.
  
 \br\label{notation} In case the set $x$ is the set of eigenvalues of a matrix $X$, we also write $s_\lambda(X)$ instead
 of $s_\lambda(x)$.
 \er

 There is a different definition of the Schur function as quasi-homogeneous non-symmetric polynomial of degree $|\lambda|$ in 
 other variables, the so-called power sums,
 $\bpow =(p_1,p_2,\dots)$, where $\deg p_m = m$.
 
For this purpose let us introduce 
$$
 s_{\{h\}}(\mathbf p)=\det[s_{(h_i+j-N)}(\mathbf p)]_{i,j},
$$
where $\{h\}$ is any set of $N$ integers, and where
the Schur functions $s_{(i)}$ are defined by $e^{\sum_{m>0}\frac 1m p_m z^m}=\sum_{m\ge 0} s_{(i)}(\bpow) z^i$.
If we put $h_i=\lambda_i-i+N$, where $N$
is not less than the length of the partition $\lambda$, then
\begin{equation}\label{Schur-t}
 s_\lambda(\mathbf p)= s_{\{h\}}(\mathbf p).
\end{equation}

 The Schur functions defined by (\ref{Schur-x}) and by (\ref{Schur-t}) are equal,  $s_\lambda(\bpow)=s_\lambda(x)$, 
 provided the variables $\bpow$ and $x$ are related by the power sums relation
  \be
\label{t_m}
  p_m=  \sum_i x_i^m
  \ee
  
  In case the argument of $s_\lambda$ is written as a non-capital fat letter  the definition (\ref{Schur-t}),
  and we imply the definition (\ref{Schur-x}) in case the argument is not fat and non-capital letter, and
  in case the argument is capital letter which denotes a matrix, then it implies the definition (\ref{Schur-x}) with $x=(x_1,\dots,x_N)$ being
  the eigenvalues.
  
  It may be easily checked that
  \be\label{p-to-p-in-Schur}
  s_\lambda(\bpow)=(-1)^{|\lambda|}s_{\lambda^{\rm tr}}(-\bpow)
  \ee
  where $\lambda^{\rm tr}$ is the partition conjugated to $\lambda$ (in \cite{Mac} it is denoted by $\lambda^*$). The Young diagram
  of the conjugated partition is obtained by the transposition of the Young diagram of $\lambda$ with respect to its main diagonal. 
  One gets $\lambda_1=\ell(\lambda^{\rm tr})$.

  \section{Integrals over the unitary group.}
 Consider the following integral over the unitary group which depends on two semi-infinite sets of parameters
 $\bpow=(p_1,p_2,\dots )$ and $\bbpow=(p_1^*,p_2^*,\dots ) $: 
  \be
I_{\mathbb{U}(N)}(\bpow,\bbpow):= \int_{\mathbb{U}(N)} 
e^{\tr V\left(\bpow ,U\right) + \tr V\left(\bpow^*,U^{-1}\right)}  d_*U=
 \ee
 \be
\frac{1}{(2\pi )^N} 
\int_{0 \le \theta_1 \le  \dots \le \theta_N\le 2\pi} 
\prod_{1\le j<k\le N}\vert e^{i\theta_j}-e^{-i\theta_k} \vert ^2 
 \prod_{j=1}^N e^{\sum_{m>0}\frac 1m \left(p_me^{im\theta_j} +p_m^* e^{-im\theta_j}\right)}d\theta_j
 \ee
 \be\label{V}
 V(\bpow,x):= \sum_{n>0} \frac 1n p_n x^n
 \ee  
 Here $d_*U$ is the Haar measure of the group $\mathbb{U}(N)$:
\be\label{Haar-unitary}
 d_*U =\frac{1}{(2\pi )^N}  
\prod_{1\le j<k\le N}\vert e^{i\theta_j}-e^{-i\theta_k} \vert ^2 
 \prod_{j=1}^N d\theta_j\,,\quad -\pi \le \theta_1<\dots\theta_N\le \pi
 \ee
 and
 $e^{i\theta_1},\dots,e^{i\theta_N }$ are the eigenvalues of $U\in  \mathbb{U}(N)$. The exponential factors
 inside the integral may be treated as a perturbation of the Haar measure and parameters $\bpow, \, \bpow^*$
 are called coupling constants by the analogy with quantum field theory problems.

 Using  the Cauchy-Littlewood identity
 \be
\label{CL}
  \tau(\bpow|\bpow^*):=e^{\sum_{m=1}^\infty \frac 1m p_m^*p_m}=\sum_{\lambda\in \mathbb{P}} s_\lambda(\bpow^*)s_\lambda(\bpow)
  \ee
 and the orthogonality of the irreducible characters of the unitary group
  \be\label{orthonormality-ch-U}
  \int s_\lambda(U)s_\mu(U^{-1})d_*U = \delta_{\lambda,\mu}
  \ee
 we obtain that
 \be\label{Morozov}
 I_{\mathbb{U}(n)}(\bpow,\bbpow) = \sum_{\lambda\in\mathbb{P}\atop
 \ell(\lambda)\le n} s_\lambda(\bpow) s_\lambda(\bbpow)
 \ee
which express the integral over unitary matrices as the "perturbation series in coupling constants".

The formula (\ref{Morozov}) first appeared in \cite{MirMorSem} in the context of the study of Brezin-Gross-Witten model.
It was shown there that the integral $ I_{\mathbb{U}(n)}(\bpow,\bbpow)$ may be related to 
the Toda lattice tau function
of \cite{JM} and \cite{UT} under certain restriction. Then, the series in the Schur functions (\ref{Morozov}) may be related 
to the double Schur functions series found in \cite{Takasaki} and \cite{Takebe}.

\section{Hurwitz numbers\label{Hurwitz-numbers} \cite{NO-LMP}}

\subsection{Definitions and examples}

For a partition $\Delta$ of a number $d=|\Delta|$ denote by $\ell(\Delta)$ the number of the non-vanishing parts.
For the Young diagram corresponding to $\Delta$, the number $|\Delta|$ is the weight  of the diagram  and $\ell(\Delta)$
is the number of rows. Denote by $(d_1,\dots,d_{\ell})$ the Young diagram with rows of length $d_1,\dots,d_{\ell}$ and
corresponding partition of $d=\sum d_i$. We need the notion of the colength of a partition $\Delta$ which is
$\ell^*(\Delta):=|\Delta|-\ell(\Delta)$.

Let us consider a connected compact surface without boundary $\Omega$ and a branched covering $f:\Sigma\rightarrow\Omega$
by a connected or non-connected surface $\Sigma$. We will consider a covering $f$ of the degree $d$. It means that the
preimage $f^{-1}(z)$ consists of $d$ points $z\in\Omega$ except some finite number of points. This points are called
\textit{critical values of $f$}.

Consider the preimage $f^{-1}(z)=\{p_1,\dots,p_{\ell}\}$ of $z\in\Omega$. Denote by $d_i$ the degree of $f$ at $p_i$. It
means that in the neighborhood of $p_i$ the function $f$ is homeomorphic to $x\mapsto x^{d_i}$. The set $(d_1\dots,d_{\ell})$
is the partition of $d$, that is called \textit{topological type of $z$}.

Fix now points $z_1,\dots,z_{\textsc{f}}$ and partitions $\Delta^{(1)},\dots,\Delta^{(\textsc{f})}$ of $d$. Denote by
\[\widetilde{C}_{\Omega (z_1\dots,z_{\textsc{f}})} (d;\Delta^{(1)},\dots,\Delta^{(\textsc{f})})\]
the set of all branched covering $f:\Sigma\rightarrow\Omega$ with critical points $z_1,\dots,z_{\textsc{f}}$ of topological
types  $\Delta^{(1)},\dots,\Delta^{(\textsc{f})}$.

Coverings $f_1:\Sigma_1\rightarrow\Omega$ and $f_2:\Sigma_2\rightarrow\Omega$ are called isomorphic if there exists an
homeomorphism $\varphi:\Sigma_1\rightarrow\Sigma_2$ such that $f_1=f_2\varphi$. Denote by $\texttt{Aut}(f)$  the group of
automorphisms of the covering $f$. Isomorphic coverings have isomorphic groups of automorphisms of degree $|\texttt{Aut}(f)|$.

Consider now the set $C_{\Omega (z_1\dots,z_{\textsc{f}})} (d;\Delta^{(1)},\dots,\Delta^{(\textsc{f})})$ of isomorphic classes
in $\widetilde{C}_{\Omega (z_1\dots,z_{\textsc{f}})} (d;\Delta^{(1)},\dots,\Delta^{(\textsc{f})})$. This is a finite set.
The sum
\[
H^{\textsc{e},\textsc{f}}(d;\Delta^{(1)},\dots,\Delta^{(\textsc{f})})=
\sum\limits_{f\in C_{\Omega (z_1\dots,z_{\textsc{f}})}(d;\Delta^{(1)},\dots,
\Delta^{(\textsc{f})})}\frac{1} {|\texttt{Aut}(f)|}\quad,
\]
don't depend on the location of the points $z_1\dots,z_{\textsc{f}}$ and is called \textit{Hurwitz number}.
Here $\textsc{f}$ denotes the number of the branch points, and $\textsc{e}$ is the Euler characteristic of the base surface.

\vspace{1ex}
{\bf Example}.
Let $f:\Sigma\rightarrow\mathbb{RP}^2$ be a covering without critical points.
Then, if $\Sigma$ is connected, then $\Sigma=\mathbb{RP}^2$,
$\deg f=1$\quad or $\Sigma=S^2$, $\deg f=2$. Therefore if $d=3$, then
$\Sigma=\mathbb{RP}^2\coprod\mathbb{RP}^2\coprod\mathbb{RP}^2$ or $\Sigma=\mathbb{RP}^2\coprod S^2$.
Thus $H^{1,0}(3)=\frac{1}{3!}+\frac{1}{2!}=\frac{2}{3}$.

\vspace{1ex}

The Hurwitz numbers arise in different fields of mathematics: from algebraic geometry to integrable systems. They are well
studied for orientable $\Omega$. In this case the Hurwitz number coincides with the weighted number of holomorphic branched
coverings of a Riemann surface $\Omega$ by other Riemann surfaces, having critical points $z_1,\dots,z_\textsc{f}\in\Omega$ of
the topological types $\Delta^{(1)},\dots,\Delta^{(\textsc{f})}$ respectively. The well known isomorphism between Riemann
surfaces and complex algebraic curves gives the interpretation of the Hurwitz numbers as the numbers of morphisms of
complex algebraic curves.

Similarly, the Hurwitz number for a non-orientable surface $\Omega$ coincides with the weighted number of the dianalytic
branched coverings of the Klein surface without boundary by another Klein surface and coincides with the weighted number
of morphisms of real algebraic curves without real points \cite{AG,N90,N2004}. An extension of the theory to all Klein surfaces
and all real algebraic curves leads to Hurwitz numbers for surfaces
with boundaries may be found in \cite{AN,N}.

\vspace{2ex}

The Hurwitz numbers have a purely algebraic description.  Any branched covering $f:\Sigma\rightarrow\Omega$ with
critical points $z_1,\dots,z_\textsc{f}\in\Omega$ generates a homomorphism $
\phi: \pi_1(u,\Omega\setminus\{z_1,\dots z_F\})\rightarrow S_{\Gamma}$,
where $u$ is a point in $\Omega$,
to the group of permutations of the set $\Gamma=f^{-1}(u)$ by the monodromy along contours of
$\pi_1(u,\Omega\setminus\{z_1,\dots z_F\})$.  Moreover, if $l_i\in\pi_1(u,\Omega\setminus\{z_1,\dots z_F\})$ is a contour
around $z_i$, then the cyclic type of the permutation $\phi(l_i)$ is $\Delta^{(i)}$. Denote by
$$\texttt{Hom}_\Omega(d;\Delta^{(1)},\dots,\Delta^{(\textsc{f})}),$$
the group of all homomorphisms $\phi: \pi_1(u,\Omega\setminus\{z_1,\dots z_F\})\rightarrow S_{\Gamma}\cong S_d$ with this
property. Isomorphic coverings generate elements of $
\texttt{Hom}_\Omega(d;\Delta^{(1)},\dots,\Delta^{(\textsc{f})})$ conjugated by $S_d$. Thus
we construct the one-to-one correspondence between $
C_{\Omega (z_1\dots,z_{\textsc{f}})} (d;\Delta^{(1)},\dots,\Delta^{(\textsc{f})})$ and the conjugated classes
of $\texttt{Hom}_\Omega(d;\Delta^{(1)},\dots,\Delta^{(\textsc{f})})$.

\vspace{1ex}

Consider the last set in more details. Any $s\in S_d$ generates the interior automorphism $I_s(g)=sgs^{-1}$ of $S_d$. Therefore
$S_d$ acts on $\texttt{Hom}_\Omega(d;\Delta^{(1)},\dots,\Delta^{(\textsc{f})})$ by $s(h)=I_s h$. The orbit of this action of
$I=\{I_s\}$ corresponds to an equivalent class of coverings. Moreover, the group $\texttt{A}= \{s\in S_d|s(h)=h\}$ is
 isomorphic to the group
$\texttt{Aut}(f)$, there the covering $f$ corresponds to the homomorphism $h$.

Consider the splitting
$\texttt{Hom}_\Omega(d;\Delta^{(1)},\dots,\Delta^{(\textsc{f})})= \bigcup\limits_{i=1}^rH_i$ on obits by $I$. Then the
cardinality $|H_i|$ is $\frac{d!}{|\texttt{A}(h_i)|}= \frac{d!}{|\texttt{Aut}(f_i)|}$, where $h_i\in H_i$. On the other
hand, the orbits $H_i$ are in the one-to-one correspondence with the classes of the coverings.
Therefore $\frac{1}{d!}|\texttt{Hom}_\Omega(d;\Delta^{(1)}, \dots,\Delta^{(\textsc{f})})| =
\frac{1}{d!}\sum\limits_{i=1}^r |H_i|= \sum\limits_{i=1}^r \frac{1}{|\texttt{Aut}(f_i)|}$ is the
Hurwitz number $H_\Omega(d; \Delta^{(1)}\,\dots,\Delta^{(\textsc{f})})$.

\vspace{2ex}

Find now $|\texttt{Hom}_\Omega(d;\Delta^{(1)},\dots,\Delta^{(\textsc{f})})|$ in terms of the characters of $S_d$. Recall, that
cyclic type  of $s\in S_d$ is cardinalities $\Delta=(d_1,\dots,d_{\ell})$ of subsets, on which the permutation $s$ split
the set $\{1,\dots,d\}$. Any partition $\Delta$ of $d$ generates the set $C_{\Delta}\subset S_d$,  consisting of permutations
of cyclic type $\Delta$. The cardinality of $C_\Delta$ is equal to
\be\label{C-Delta,z-Delta}
|C_\Delta| \,=\,\frac{|\Delta|!}{z_\Delta}\,,\qquad
z_\Delta\,=\,\prod_{i=1}^\infty \,i^{m_i}\,m_i!
\ee
where $m_i$ denotes the number of parts equal to $i$ of the partition $\Delta$ (then the partition $\Delta$ is often
denoted by $(1^{m_1}2^{m_2}\cdots)$).

Moreover, if $s_1,s_2\in C_{\Delta}$, then $\chi(s_1)=\chi(s_2)$ for any complex character $\chi$ of $S_d$. Thus we can
define $\chi(\Delta)$ for a partition $\Delta$, as $\chi(\Delta)=\chi(s)$ for $s\in C_\Delta$.

The Mednykh-Pozdnyakova-Gareth A. Jones formula \cite{M1,M2,ZL,GARETH.A.JONES} says that
\[ \big|\texttt{Hom}_\Omega(d;\Delta^{(1)},\dots, \Delta^{(\textsc{f})})\big| = \,
d!\sum_{\lambda} \,
\left(\frac{{\rm dim}\lambda}{d!}\right)^{\textsc{e}}\,
\prod_{i=1}^\textsc{f} \,\,|C_{\Delta^{(i)}}|\,\,
\frac{\chi_\lambda(\Delta^{(i)})}
{{\rm dim}\lambda} \,,
\]
where $\textsc{e}=\textsc{e}(\Omega)$ is the Euler characteristic of $\Omega$ and $\chi_\lambda$ ranges over the irreducible
complex characters of
$S_d$, associated with Young diagrams of weight $d$. Thus we obtain
\be\label{Hurwitz-counting}
H^{\textsc{e},k}(d;\Delta^1,\dots,\Delta^{k})=\sum_{\lambda\atop |\lambda|=d}
\left(\frac{{\rm dim}\lambda}{d!} \right)^{\textsc{e}}\varphi_\lambda(\Delta^1)\cdots 
\varphi_\lambda(\Delta^k)
\ee

In order to get Hurwitz numbers for the projective plane $\mathbb{RP}^2$ we choose $\textsc{e}=1$, and for the Riemann sphere we choose $\textsc{e}=2$.

\vspace{1ex}

{\bf Example}. Let $\textsc{e}=1$, $\textsc{f}=0$, $d=3$.
Then, $ H^{1,0}(3) = \sum_{|\lambda|=3}
\frac{{\rm dim}\lambda}{d!}=\frac{4}{6}=\frac{2}{3}$.
In general for the unbranched covering  of $\mathbb{RP}^2$ we get the following generating function
(compare to (\ref{r=1}) below)
\be\label{unbranched}
e^{\frac {c^2}{2} + c}\,=\,\sum_{d\ge 0}\, c^d H^{1,0}(d)
\ee
The exponent reflects the fact that the connected unbranched covers of the projective plane may consist of either
the projective plane (the term $c$: 1-fold cover) or the Riemann sphere - opposite points of the sphere correspond 
to a point of the projective plane (the term $\frac{c^2}{2}$: 2-fold cover,  where $2$ in the
denominator is the order of the automorphisms of the covering by the sphere).

At the end we write down purely combinatorial definition of the projective Hurwitz numbers \cite{M2}, \cite{GARETH.A.JONES}.
Let us consider the symmetric group $S_d$ and  the equation
\be\label{equation-in-S-d}
R^2 X_1 \cdots X_{\textsc{f}} = 1,\quad R,X_i \in S_d,\quad X_i \in C_{\Delta^{(i)}},\, i=1,\dots,\textsc{f}
\ee
where $C_{\Delta^{(i)}},\, i=1,\dots,\textsc{f}$ are the cycle classes of a given set of partitions
$\Delta^{(i)},\,i=1,\dots,\textsc{f}$ of a given weight $d$. Then
$H^{1,\textsc{f}}(d;\Delta^{(1)},\dots,\Delta^{(\textsc{f})})$ is the number of solutions to (\ref{equation-in-S-d}) divided
over $d!$. Say, for unbranched 3-fold covering we get 4 solutions to the  equation $R^2=1$ in the permutation group $S_3$: 
the unity element and three transpositions.
Thus $H^{1,0}=4:3!$ as it was obtained in the Examples above. The number of solutions to the equation $R^2=1$ in $S_d$ is given by
(\ref{unbranched}).

\subsection{Remarks on Mednykh-Pozdnyakova-Gareth A. Jones character formula\label{Remarks-on-Mednykh-subsection} \cite{NO-LMP}}

Let's begin with the remark concerning the context of this subsection:

\br\label{Natanzon}

It follows from R.Dijkgraaf paper \cite{Dijkgraaf} that the Hurwitz numbers for closed orientable surfaces form 2D topological field theory.
An extension of this result to the case of Klein surfaces was found in \cite{AN},
theorem 5.2. (see also Corollary 3.2 in \cite{AN2008}) On the other hand Mednykh-Pozdnyakova-Gareth A. Jones formula gives the description of
the Hurwitz numbers in terms of characters of the symmetric groups. In this Subsection in fact we give the interpretation  of axioms of
the Klein topological field theory \cite{AN} for Hurwitz numbers in terms of characters of symmetric groups, this approach is different
from \cite{AN}.

\er

(A) Let us present the following statement
\bl
\label{Hurwitz-down-Lemma}
\be\label{Hurwitz=Hurwirz-Hurwitz}
H^{\textsc{e}+\textsc{e}_1,\textsc{f}+\textsc{f}_1}(d;\Delta^{(1)},\dots,\Delta^{(\textsc{f}+\textsc{f}_1)})=
\sum_{\Delta}\frac{d!}{|C_\Delta|}
H^{\textsc{e}+1,\textsc{f}+1}(d;\Delta^{(1)},\dots,\Delta^{(\textsc{f})},\Delta)
H^{\textsc{e}_1+1,\textsc{f}_1+1}(d;\Delta,\Delta^{(\textsc{f}+1)},\dots,\Delta^{(\textsc{f}_1)})
\ee
In particular
\be\label{Hurwitz-down}
H^{\textsc{e}-1,\textsc{f}}(d;\Delta^{(1)},\dots,\Delta^{(\textsc{f})})=
\sum_{\Delta}\,
H^{\textsc{e},\textsc{f}+1}(d;\Delta^{(1)},\dots,\Delta^{(\textsc{f})},\Delta) \chi(\Delta)
\ee
where $\chi(\Delta)=\frac{d!H^{1,1}(d;\Delta)}{|C_\Delta|}$ are rational numbers explicitly defined by a partition $\Delta$ as follows
\be\label{delta(Delta)}
\chi(\Delta)=\sum_{\lambda \atop |\lambda|=|\Delta|} \chi_\lambda(\Delta)=\left[
\prod_{i>0,\,{\rm even}}e^{\frac {i}{2}\frac{\partial^2}{\partial p_i^2}}\cdot p_i^{m_i}
\prod_{i>0,\,{\rm odd}}e^{\frac {i}{2}\frac{\partial^2}{\partial p_i^2}+
\frac{\partial}{\partial p_i}}\cdot p_i^{m_i}
\right]_{\bpow =0}
\ee
where $\chi_\lambda(\Delta)$ is the character of the representation $\lambda$ of the symmetric group
$S_d$, $d=|\lambda|$, evaluated on the cycle class $\Delta=(1^{m_1}2^{m_2}\cdots)$.

\el
In particular, we get that the Hurwitz numbers of the projective plane may be obtained from
the Hurwitz numbers of the Riemann sphere while the Hurwitz numbers of the torus and of the Klein bottle
may be obtained from Hurwitz numbers of the projective plane.

First we prove the second equality in (\ref{delta(Delta)}). It
follows from the relations
\be\label{Laplace=sum-Schurs}
e^{\sum_{i>0}\frac {i}{2}\frac{\partial^2}{\partial p_i^2}+
\sum_{i>0,\,{\rm odd}}\frac{\partial}{\partial p_i}}=\sum_{\lambda} s_\lambda({\tilde\partial})
\ee
\be\label{Schur-orthogonality}
 \left[s_\lambda({\tilde\partial})\cdot s_\mu(\bpow)   \right]_{\bpow=0}=\delta_{\lambda,\mu}\,,
\qquad
p_1^{m_1}p_2^{m_2}\cdots =:p_\Delta =\sum_{\lambda} \chi_\lambda(\Delta) s_\lambda(\bpow)
\ee
where $s_\lambda({\tilde\partial})$ is the Schur function $s_\lambda(\bpow)$ where each $p_i$ in its argument is replaced by
the derivative $i\frac{\partial}{\partial p_i}$. The heat operator in the left hand side of (\ref{Laplace=sum-Schurs}) plays an important role.
The relations (\ref{Schur-orthogonality})
may be found in \cite{Mac}. The relation (\ref{Laplace=sum-Schurs}) is derived from the known relation
\be\label{sum-Schurs}
 \sum_\lambda s_\lambda(\bpow({\bf x}))=\prod_{i<j}\frac{1}{1-x_ix_j}\prod_{i}\frac{1}{1-x_i},
 \quad p_m({\bf x}):=\sum_i x_i^m
\ee
which also may be found in \cite{Mac}.

Equality (\ref{Hurwitz=Hurwirz-Hurwitz}) follows from the orthogonality relation for characters:
$\sum_\Delta |C_\Delta| \chi_\lambda(\Delta)\chi_\mu(\Delta) = d!\delta_{\mu,\lambda}$ where $|\mu|=|\lambda|=|\Delta|=d$
which yields $\sum_\Delta \varphi_\lambda(\Delta)\chi(\Delta)=d!\left({\rm dim}(\lambda)\right)^{-1} $.  Then the formula
(\ref{Hurwitz-counting})  gives (\ref{Hurwitz=Hurwirz-Hurwitz}).

In (\ref{2KP-to-BKP}) below we shall see that the heat operator which enters formula (\ref{delta(Delta)}) also links solutions of 2KP (TL)
and BKP hierarchies.

(B) Another remark is as follows.

Let us use the so-called Frobenius notation \cite{Mac} for a partition $\lambda$: $\lambda=(\alpha_1,\dots,\alpha_{\kappa}|\beta_1,\dots,\beta_\kappa)$,
$\alpha_1>\cdots >\alpha_\kappa\ge 0$, $\beta_1> \cdots >\beta_\kappa \ge 0$. The integer $\kappa=\kappa(\lambda)$ denotes the length
of the main diagonal of the Young diagram $\lambda$, the length of $\lambda$ is denoted by $\ell(\lambda)$.

\bl\label{character-on-d-cycle-lemma}
The normalized character labeled by $\lambda$ evaluated at cycle class $\Delta$ (as usual $d=|\lambda|$) vanishes if $\kappa(\lambda)>1$, 
and may explicitly evaluated at the cycle $\Delta=(d)$:
\be\label{character-on-d-cycle}
\varphi_\lambda\left((d)\right) = (-1)^{\ell(\lambda)+1}\left( \frac{d!}{{\rm dim}\lambda}  \right)\frac 1d \delta_{1,\kappa(\lambda)}
\ee
\el
For the proof we at first will notice that the Schur function marked with a Young diagram consisting of one hook,
say, $ (\alpha_i |\beta_j)$ may be written in the following form:
\[
 s_{(\alpha_i|\beta_j)}(\bpow) = \frac 1d (-1)^{\beta_j}p_{\alpha_i+\beta_j+1} + \cdots
\]
where dots denote terms which do not depend on $p_a$, $a\ge \alpha_i+\beta_j+1$ (this fact may be derived, say, from the Jacobi-Trudi
formula $s_\lambda(\bpow)=\det\, s_{(\lambda_i-i+j)}(\bpow)$). Then from the Giambelli identity
$s_\lambda(\bpow)=\det\,s_{(\alpha_i|\beta_j)}(\bpow)$ it follows that $s_\lambda$ does not depend on $p_a$, $a>\alpha_1+\beta_1+1$. Thus
it does not depend on $p_a$, $a\ge d>\alpha_1+\beta_1 +1$ in case $\kappa(\lambda)>1$. Due to the character map relation it means that
$\varphi_\lambda\left((d)\right)=0$ if $\kappa(\lambda)>1$. In case of a one-hook partition $\lambda=(\alpha_1|\beta_1)$ we have
$\alpha_1+\beta_1+1 =d$ and the character
map formula (\ref{Schur-char-map}) yields
\[
 s_{(\alpha_1|\beta_1)}(\bpow)=\frac{{\rm dim}\lambda}{d!}\left(p_d\varphi_\lambda\left((d)\right) +\cdots \right)
\]
where dots denote terms which does not depend on $p_d$. We compare last two formulae and get (\ref{character-on-d-cycle}).

For $d$-fold covers we shall call a branch point \textit{maximally ramified} in case its ramification profile is $(d)$.
It describes merge in this branch point of all $d$ sheets of the covering surface.

Formula (\ref{character-on-d-cycle}) allows to equate Hurwitz numbers related to different
Euler characteristics of
base Klein surfaces if in both
cases there are nonvanishing numbers of ramification profiles $(d)$. Namely,
 the Mednykh-Pozdnyakova-Gareth A. Jones character formula (\ref{Hurwitz-counting}) yields
\bp\label{d-cycle-proposition} For any natural number $\textsl{g}$
\be\label{H-H}
H^{\textsc{e}-2\textsl{g}, \textsc{f}+1}\left(d;\Delta^{(1)},\dots,\Delta^{(\textsc{f})},(d)  \right)=
d^{2\textsl{g}}
H^{\textsc{e}, \textsc{f}+2\textsl{g}+1}\left(d;\Delta^{(1)},\dots,\Delta^{(\textsc{f})},(d),
\underbrace{(d),\dots,(d)}_{2\textsl{g}}  \right)
\ee
\ep
For the first time this formula was proven by Zagier for the case of even $\textsc{e}$, see Appendix A in the wonderful book \cite{ZL}. 
We get it in a different way and for all $\textsc{e}$.

\br\label{connected-via-(d)-profile}
Notice that the presence of the profile $(d)$ means that the cover is connected.
\er

\br
\label{Zagier}
In the Appendix A by Zagier in \cite{ZL} the polynomial
$\texttt{R}_\Delta(\texttt{q}):=\frac{\prod_{i=1}^{\ell{\Delta}}(1-\texttt{q}^{d_i})}{1-\texttt{q}}=
:\sum_r (-1)^r\texttt{q}^r \chi_r(\Delta) $
was introduced. It was shown that $\chi_r $ ($0\le r \le d-1 $) is the character of the irreducible representation of $S_d$ given
by $\chi_r(g)=\tr (g,\pi_r)$, $g\in S_d$,
$\pi_r={\wedge}^r({\rm St}_d) $. Here ${\rm St}_d$ is the vector space $\{ (x_1,\dots,x_d)\in \mathbb{C}^d | x_1+\cdots+x_d=0 \}$ and
$S_d$ acts by permutations of the coordinates.
We can make a specification: the character $\chi_r$ coincides with $\chi_\lambda$ where $\lambda=(d-r|r)$. 
That to understand it, we will consider the Schur function
$s_\lambda(\bpow(\texttt{q},0))$ where $p_m(\texttt{q},0):=1-\texttt{q}^m$ and $\lambda$ is not yet fixed. There is the 
following equation 
$$
s_\lambda(\bpow)=(-1)^{\ell(\lambda)-1} (1-\texttt{q})\texttt{q}^{\ell(\lambda)-1} \delta_{\kappa(\lambda),1}\, ,
$$
for instance, see Ex 23, I.3 in \cite{Mac}. In the this relation
denote $k=\ell(\lambda)-1$, then $\lambda=(d-k|k)$.
On the other hand, formula (\ref{Schur-char-map}) says 
$$
s_{(d-k|k)}(\bpow(\texttt{q},0))=(1-\texttt{q})\sum_{\Delta} \frac{|C_\Delta |}{d!} \chi_{(d-k|k)}(\Delta) \texttt{R}_\Delta(\texttt{q})=
(1-\texttt{q})\sum_{r=0}^{d-1}(-\texttt{q})^r \sum_\Delta \frac{|C_\Delta|}{d!}\chi_{(d-k|k)}(\Delta)\chi_r(\Delta)
$$
Compare last relations.
The orthogonality of characters leads to equality $\chi_k = \chi_{(d-k|k)}$. It, by the way, means that in the presence of a 
maximally ramified branch point the summation range in (\ref{Hurwitz-number}) is limited to one-hook Young diagrams $\lambda$.

\er
Formula (\ref{tau-Hurwitz-themselves}) will show below how tau functions generate
 Hurwitz numbers of covers with the maximally ramified branched points.

\section{Two-component fermions, two-component KP tau functions and multi-matrix models \label{Product-of-matrices-and-2KP}}

The development of the topic below was planned in \cite{OrlovStrahov}.

\subsection{Fermions \label{fermions} }

Introduce Fermi operators 
$\psi_i^{(\alpha)}$ and $\psi_i^{\dag(\alpha)}$ where $i\in\mathbb{Z}$ 
and $\alpha=1,\dots, p$ with the following
canonical anticommutation relations
\be\label{anticommutation}
\psi_i^{(\alpha)}\psi_j^{\dag (\beta)} +\psi_j^{\dag(\beta)}  \psi_i^{(\alpha)}
=\delta_{i,j}\delta_{\alpha,\beta}\,,\quad \alpha=1,\dots,p
\ee
and all other anticommutators vanish. Below we shall use only the cases $p=1,2$ (namely, one- and two-component fermions).
In case $p=1$ we shall omit the superscript: $\psi_i:=\psi^{(1)}$, $\psi_i^\dag:=\psi_i^{\dag(1)}$.

Left and right vacuum vectors $\langle 0;0|$ and $|0;0\rangle$  ($\langle 0;0|1|0;0\rangle =1$):
\be\label{annihilation}
\langle 0;0|\psi_{-i-1}^{(\alpha)}=\langle 0;0|\psi_i^{\dag(\alpha)}=0=
\psi_i^{(\alpha)} |0;0\rangle = \psi_{-i-1}^{\dag(\alpha)} |0;0\rangle\,,
\quad i<0\,,\quad \alpha=1,2
\ee

Given partitions $\lambda=(\lambda_1,\dots,\lambda_N)$, 
$\nu=(\nu_1,\dots,\nu_N)$ introduce the following Fock vector
\be\label{right-Fock}
|N,\lambda;-N,\nu \rangle =
\psi_{\nu_1-1+N}^{(2)}\cdots \psi_{\nu_N}^{(2)}\psi_{\lambda_1-1+N}^{(1)}\cdots \psi_{\lambda_N}^{(1)}|0;0\rangle
\ee

\subsection{Tau functions of the two-component KP \label{tau-functions-2KP-section}  \cite{JM}}

It is known \cite{JM} that the following expression
\be\label{Fermi-expectation}
\mathbb{I}_N(p,p^*)=\langle N;-N| \Gamma^{(1)}(p)\Gamma^{(2)}(p^*) g |0;0\rangle
\ee
with
\be\label{g-2KP,TL}
g = \exp\, \sum_{\alpha,\beta=1,2}\sum_{i,j\in\mathbb{Z}} 
\psi_i^{(\alpha)}\psi_j^{\dag(\beta)} C_{i,j}^{\alpha,\beta}
\ee
\be\label{Gamma-1-2}
\Gamma^{(\alpha)}(p)=\sum_{k>0} \frac 1k p_k J_k^{(\alpha)}\,,\quad 
J_k^{(\alpha)}=\sum_{i\in\mathbb{Z}} \psi_i^{(\alpha)}\psi_{i+k}^{\dag(\alpha)}
\ee
is an example of the two-component KP tau function and solves Hirota equations
for the 2-component KP (and also an example of the semi-infinite Toda 
lattice tau function). Infinite sets $p=(p_1,p_2,\dots)$ and $p^*=(p_1^*,p_2^*,\dots)$ are called
higher times of the 2-KP. Constants $C_{i,j}^{\alpha,\beta}$ play the role the integrals of motion
which define the solution of the nonlinear equations resulting from the Hirota equations and equivalent to them.

Now let us introduce Fermi fields
\be\label{Fermi-field}
\psi^{(\alpha)}(z):=\sum_{i\in\mathbb{Z}}\, \psi_i^{(\alpha)} z^i\,,
\quad
\psi^{\dag(\alpha)}(z) :=\sum_{i\in\mathbb{Z}}\, \psi_i^{\dag(\alpha)} z^{-i-1}
\ee
We have 
\be\label{Gamma-vac}
\Gamma^{(\alpha)} |0;0\rangle = |0;0\rangle
\ee
which follows from $J^{(\alpha)}_k|0;0\rangle =0$.
We need  the known equations
\be\label{p-dependence}
\Gamma^{(\alpha)}(p)\psi^{(\alpha)}(z)=e^{V(z,p)}\psi^{(\alpha)}(z)\Gamma^{(\alpha)}(p)\,,
\quad 
\Gamma^{(\alpha)}(p)\psi^{\dag(\alpha)}(z)=e^{-V(z,p)}\psi^{\dag(\alpha)}(z)\Gamma^{(\alpha)}(p)
\ee
which may be obtained from the relations $[J_k^{(\alpha)},\psi_i^{(\alpha)}]=\psi_{i-k}^{(\alpha)}$ and 
$[J_k^{(\alpha)},\psi_i^{\dag(\alpha)}]=\psi_{i+k}^{\dag(\alpha)}$ following from (\ref{anticommutation}).

We need in the known relation
\be\label{Schur-via-fermions}
\langle N,-N|\Gamma^{(1)}(p)\Gamma^{(2)}(p^*) |N,\lambda ;-N,\nu \rangle = s_\lambda(p) s_\nu(p^*)
\ee
where $s_\lambda$ are the Schur functions.

\subsection{Multiple integrals \cite{HO-2006},\cite{NO-2014} \label{multi-integrals-section}}

Let us specify (\ref{g-2KP,TL}) as follows
\be\label{g-for-MM}
g=\exp\,c\int_{\mathbb{C}} \psi^{(1)}(z) \psi^{\dag(2)}({\bar z}) \mu(z,{\bar z})d^2 z
\ee
We have
\bl\label{Vandermond}
\be\label{Vandermonds}
\langle N;-N |  \psi^{(1)}(z_1)\cdots   \psi^{(1)}(z_N) \psi^{\dag(2)}({\bar z}_1)\cdots \psi^{\dag(2)}({\bar z}_N)
|0;0\rangle= (-1)^{...}|\Delta(z)|^2
\ee
\el
For the proof we insert series (\ref{Fermi-field}) into the left hand side of (\ref{Vandermonds}) and 
use (\ref{anticommutation}) and (\ref{annihilation}).

From Lemma \ref{Vandermond} and from (\ref{p-dependence}) and (\ref{Gamma-vac}) we get
\be
\mathbb{I}_N(p,p^*)=\frac{c^N}{N!}\int |\Delta(z)|^2 \prod_{i=1}^N \mu(z_i,{\bar z}_i) e^{V(z_i,p)-V({\bar z}_i,p^*)} d^2 z_i
\ee

Thus the right hand side is an example of the two-component KP tau function.

Next we insert (\ref{Fermi-field}) into (\ref{g-for-MM}) and obtain
\be\label{g-Fourier}
g=\exp c\sum_{i,j} \psi_i^{(1)}\psi_j^{(2)} m_{i,j},\quad m_{i,j}=\int z^i{\bar z}^j \mu(z,{\bar z}) d^2 z
\ee
 From (\ref{Fermi-expectation}) it follows
\be
g|0;0\rangle = \sum_{N\ge 0}\frac{c^N}{N!}\sum_{\lambda,\mu} |N,\lambda;-N\nu \rangle  g_{\lambda,\nu}(N)
\ee
where
\be\label{g-det}
g_{\lambda,\nu}(N)=\det \left( m_{h_j,h'_j}\right)|_{i,j=1,\dots,N}\,\quad h_i=\lambda_i-i+N\,,\quad 
h'_i=\nu_i-i+N
\ee
We imply that $g_{\lambda,\nu}(N)$ vanishes if the length of any of $\lambda ,\nu$ exceeds$N$.

And from (\ref{Schur-via-fermions}) we get
\be\label{Schur-series-general}
\mathbb{I}_N(p,p^*)=\frac{c^N}{N!}\sum_{\lambda,\nu\atop \ell(\lambda),\ell(\nu)\le N} s_\lambda(p) s_\nu(p^*) g_{\lambda,\nu}(N)
\ee

\subsection{Hypergeometric functions as tau functions of the two-component KP \label{Hyper-func-as-tau-2KP-section}}

In \cite{OS-TMP} it was shown that the so-called hypergeometric functions of matrix arguments 
\be\label{hypergeometric-function}
_nF_q\left(p,p^*\biggl|\begin{array}{cccc}
a_1+N, & a_2+N, & \ldots, & a_n+N  \\ b_1+N, & b_2+N, & \ldots, & b_q+N 
\end{array}
\right) 
\ee
$$
=\sum_{\lambda} s_\lambda(p)s_\lambda(p^*)
\frac{\prod_{\alpha =1}^n (a_\alpha+N)_\lambda}{\prod_{\alpha=1}^q (b_\alpha+N)_\lambda}=\tau^{\rm TL}(N,p,p^*)
$$
is the example of the TL tau function\footnote{By matrix arguments one implies matrices, say, $A$ and $B$ such that 
$p_k=\Tr A^k$ and $p_k^*=\Tr B^k$.}. Here $N$ is the site number (discrete time variable) and two sets 
$p=(p_1,p_2,\dots)$ and $p^*=(p_1^*,p_2^*,\dots)$ are higer times of the relativistic Toda lattice written in form
$$
\tau^{TL}(N,p,p^*) \partial_{p_1}\partial_{p_1^*}\tau^{\rm TL}(N,p,p^*)-
\partial_{p_1}\tau^{\rm TL}(N,p,p^*)\,\partial_{p_1^*}\tau^{\rm TL}(N,p,p^*)
$$
$$
= N \tau^{\rm TL}(N+1,p,p^*)\tau^{\rm TL}(N-1,p,p^*)
$$

It is well-studied and has a number of applications. 
However in genaral case it was not written in form of 2-KP tau function (\ref{Fermi-expectation}).
Let us do it now. To get the factor which follows the product of two Schur functions we choose
\begin{equation}i
\label{WeightFunction'}
\mu(z,{\bar z})=\pi^{}G_{q,n}^{n,0}\left(|z|^2\biggl|
\begin{array}{cccc}
  b_1 & b_2 & \ldots & b_q \\
  a_1 & a_2 & \ldots & a_n 
\end{array}
\right).
\end{equation}
where $G$ is the Meijer function. 
Indeed as it is well-known en.wikipedia.org-wiki-Meijer-G-function that
\begin{equation}
\label{G-n0-qn-moments}
m_{i,k}= \delta_{i,k}
\int_0^{+\infty} x^{k} G_{q,n}^{n,0}\left(x\biggl|\begin{array}{cccc}
b_1, & b_2, & \ldots, & b_q  \\ a_1, & a_2, & \ldots, & a_n 
\end{array}
\right) dx = \delta_{i,k}\frac{\prod_{\alpha=1}^n \Gamma\left(a_\alpha +k+1 \right)}{\prod_{\alpha=1}^q \Gamma\left(b_\alpha +k+1 \right)}
\end{equation}
Evaluating (\ref{g-det}) in this case we see that the series in the right hand side of (\ref{Schur-series-general}) 
result in $_{n}F_q$ of (\ref{hypergeometric-function}).

Next we show that $_{n}F_0$ coinsides with the partition function for the interacting complex Ginibre ensembles.

\subsection{Products of complex matrices and and tau function of two-component KP \label{Product-tau-2KP}}

Consider the ensemble of the interacting complex matrices given by the partition function
\be\label{n-product-ensemble}
\mathbb{I}_N^{\beta=2}(p,p^*;a_1,\dots,a_n) = \int \cdots \int 
e^{\Tr V(Z,p) + \Tr V(Z^\dag,p^*)}
\prod_{\alpha=1}^n e^{-\Tr Z_\alpha^\dag Z_\alpha} \det\left(Z^\dag Z \right)^{a_\alpha}  dZ_\alpha
\ee
where $Z=Z_1\cdots Z_n$, and $V(x,p)$ is given by (\ref{V}) and $a_\alpha$ are given numbers.
We call $p=\left(p_1,p_2,\dots \right)$ and $p^*=\left(p_1^*,p_2^*,\dots \right)$ coupling constants.

From the results of the previous sections we come to the following

{\bf Theorem}
 The partition function $\mathbb{I}_N^{\beta=2}(p,p^*;a_1\dots,a_n)$ is the tau function of the two-component KP equation given by 
 \be\label{n-product-ensemble-in-Schur}
\mathbb{I}_N^{\beta=2}(p,p^*;a_1,\dots,a_n)\,=\,{_nF}_0\left(p,p^*\biggl|\begin{array}{cccc}
a_1+N, & a_2+N, & \ldots, & a_n+N  
\end{array}
\right) 
\ee
$$
=\,\sum_{\lambda}\, s_\lambda(p)s_\lambda(p^*)\,
\prod_{\alpha =1}^n (a_\alpha+N)_\lambda
$$
 The fermionic expression for this tau function is as follows
 \be\label{n-product-ensemble-in-fermions}
\mathbb{I}_N^{\beta=2}(p,p^*;a_1,\dots,a_n)\,=\,\langle N;-N| \,\Gamma^{(1)}(p)\Gamma^{(2)}(p^*)\, e^{c\Phi_n(a_1,\dots,a_n)} \,|0;0\rangle
\ee
where
\be
\Phi_n(a_1,\dots,a_n)\,=\,\int_{\mathbb{C}} \psi^{(1)}(z) \psi^{\dag(2)}({\bar z}) w^{a_1,\dots,a_n}_n(z,{\bar z})d^2 z
\ee

{\bf Proof} follows from the previous subsections. Let us start with the relation (\ref{n-product-ensemble-in-fermions}). 
Then the fact that $\mathbb{I}_N^{\beta=2}$ is the tau function follows from
the Subsection \ref{tau-functions-2KP-section}. From the Subsection \ref{multi-integrals-section} it follows that
the fermionic expectation is equal to
\be\label{eig-integral-for-interacting-matrices}
\mathbb{I}_N^{\beta=2}(p,p^*;a_1,\dots,a_n)=
\int_{\mathbb{C}} |\Delta(z)|^2 \prod_{i=1}^N   e^{V(z_i,p)-V({\bar z}_i,p^*)} w_n^{(a_1,\ldots,a_n)}(z) d^2 z_i
\ee
{\bf Remark} One can consider a more general interacting ensemble:

\be\label{n-product-ensemble-genaral}
\mathbb{I}_N^{\beta=2}(p,p^*,L_1,L_2;a_1,\dots,a_n) = 
\ee
$$
\int \cdots \int 
e^{\Tr V(Z,p) - \Tr V(Z^\dag,p^*)}\det Z^{L_1}\det \left(Z^\dag \right)^{-L_2}
\prod_{\alpha=1}^n e^{-\Tr Z_\alpha^\dag Z_\alpha} \det\left(Z^\dag Z \right)^{a_\alpha}  dZ_\alpha
$$
which is also the tau function of the 2-KP, and it's fermionic representation is 
$$
\mathbb{I}_N^{\beta=2}(p,p^*,L_1,L_2;a_1,\dots,a_n) =
\l L_1+N,L_2-N|\,\Gamma^{(1)}(p),\Gamma^{(2)}(p^*)\, 
e^{c\Phi_n(a_1,\dots,a_n) } 
\,|L_1,L_2\r
 $$

\section{BKP tau functions. \label{BKP-tau-function}  \cite{NO-LMP} }

\subsection{BKP hierarchy of Kac and van de Leur.}

There are two different BKP hierarchies of integrable equations, one was introduced by the Kyoto group in
\cite{JM}, the other was
introduced by V. Kac and J. van de Leur in \cite{KvdLbispec}. We need the last one. This hierarchy includes
the celebrated
KP one as a particular reduction.  In a
certain way (see \cite{LeurO-2014}) the BKP hierarchy may be related to the three-component KP hierarchy
introduced in \cite{JM}
(earlier described in \cite{ZakharovShabat} with the help of L-A pairs of differential operators with matrix
valued coefficients).
For a detailed
description of the BKP we refer readers to the original work \cite{KvdLbispec}, and here
we write down the first non-trivial equation (Hirota equation) for the BKP tau function. This is
\bea\label{Hirota-elementary-1'}
\frac 12 \frac{\partial\tau(N,n,\bpow)}{\partial p_2} \tau(N+1,n+1,\bpow)-
\frac 12 \tau(N,n,\bpow)\frac{\partial\tau(N+1,n+1,\bpow)}{\partial p_2} \nonumber
+\frac 12 \frac{\partial^2\tau(N,n,\bpow)}{\partial^2 p_1} \tau(N+1,n+1,\bpow)\\
+\frac 12 \tau(N,n,\bpow)\frac{\partial^2\tau(N+1,n+1,\bpow)}{\partial^2 p_1}
- \frac{\partial\tau(N,n,\bpow)}{\partial p_1}\frac{\partial\tau(N+1,n+1,\bpow)}{\partial p_1}
=\tau(N+2,n+2,\bpow)\tau(N-1,n-1,\bpow)
\eea

The BKP tau functions depend on the set of higher continuous times $t_m=\frac 1m p_m$, $m>0$ and the discrete parameter
$N$.
In \cite{OST-I}  we added the second discrete parameter $n$, and the simplest Hirota equation relating the
BKP tau
functions for neighboring values of $n$ is
\bea\label{Hirota-elementary-2'}
\frac 12 \tau(N,n+1,\bpow)\frac{\partial^2 \tau(N+1,n+1,\bpow)}{\partial^2 p_1}-
\frac 12 \frac{\tau(N,n+1,\bpow)}{\partial^2 p_1} \tau(N+1,n+1,\bpow)=\nonumber
\\
\frac{\partial\tau(N+2,n+2,\bpow)}{\partial p_1}\tau(N-1,n,\bpow)-
 \frac{\partial \tau(N+1,n+2,\bpow)}{\partial p_1}\tau(N,n,\bpow)
\eea

 The complete set of  Hirota equations with two discrete parameters is written down
in the Appendix.

The general solution to the BKP Hirota equations may be written as
\be\label{BKP-tau-general}
\tau\left(N,n,\bpow \right)=\sum_{\lambda}\,A_\lambda(N,n) s_\lambda(\bpow)
\ee
where $A_\lambda$ satisfies the Plucker relations for an isotropic Grassmannian and (as one can show with the help of the Wick formula) may be written
in pfaffian form.

\subsection{BKP tau function of the hypergeometric type.}
We are interested in a certain subclass of the BKP tau functions (\ref{BKP-tau-general}) introduced in
\cite{OST-I} and called
BKP hypergeometric tau functions, which may be compared to a similar class of TL  and KP
tau functions found in \cite{KMMM}, \cite{OS-2000}.

Similarly to  \cite{OS-2000} we proceed as follows. Suppose that $\lambda$ is a Young diagram.
Given an arbitrary function $r$ of one variable
we construct the following product
\be
r_\lambda(x):=\prod_{(i.j)\in\lambda}\, r(x+j-i)
\ee
which is called the content product (or, sometimes, the generalized Pochhammer symbol attached to a Young
diagram $\lambda$). 

\br\label{gf}
 (1) If $r=fg$, then $r_\lambda(x)=f_\lambda(x)g_\lambda(x)$.
 (2) If ${\tilde r}(x)=\left(r(x)\right)^n,\,n\in\mathbb{C}$,
then ${\tilde r}_\lambda(x)=
\left({r}_\lambda(x)\right)^n$.
\er

We consider sums over partitions of the form
\be\label{hypergBKP-sums}
\sum_{\lambda \atop \ell(\lambda)\le N} \, r_\lambda(n)\, c^{|\lambda|}\, s_\lambda(\bpow) \,=
: \tau_r^{\rm B}(N,n,\bpow)
\ee
where $s_\lambda$ are the Schur functions and $\bpow$
denotes the semi-infinite set $(p_1,p_2,\dots )$.
It was shown in \cite{OST-I} that up to a factor (\ref{hypergBKP-sums}) defines the BKP tau function:
\bp
For any given $r$ the tau function $g(n)\tau^{\rm B}_r(N,n,\bpow)$ solves the BKP Hirota equations. Here
 $g(n)$ is a function of the parameter $n$
defined by (\ref{g(n)}) in the Appendix \ref{fermionic-appendix}.
\ep
Let us mark two points: first, though discrete parameters enter Hirota equations, for our purposes the factor $g(n)$ is
unimportant, and second, the parameter $N$ which restricts the partition length in sums over partitions should be chosen 
large enough, and we can take $N=+\infty$.

We call such tau functions hypergeometric because hypergeometric functions of one variable may be obtained as specifications of
(\ref{hypergBKP-sums}). For instance, one can choose $p_m=x^m$. Then a rational function $r$ in (\ref{hypergBKP-sums})
yields the generalized hypergeometric function of one variable while trigonometric $r$ results in the basic (the same, 
the $q$-deformed) one. However the key tau function is the simplest one:

{\bf Example}. Consider $r(x)=1 $ for any $x$. Such tau function does not depend on $n$ and will be denoted
by $\tau_1(N,\bpow)$. Other hypergeometric tau functions may be obtained by action of a specially chosen vertex operator
on $\tau_1(N,\bpow)$, for example see (\ref{tau-via-vertex-t}). If we take $N=+\infty$ we can obtain
\be\label{r=1}
\tau^{\rm B}_1(\infty,\bpow)=\sum_\lambda c^{|\lambda|} s_\lambda(\bpow)=
e^{\sum_{m>0} \left(\frac {c^2}{2m}p_m^2 +c\frac {p_{2m-1}}{2m-1}\right)}
\ee

\br
Each tau function $\tau_r^{\rm B}$ may be expressed as a pfaffian, see \cite{OST-I}.
\er

\paragraph{As are connected 2-KP and BKP hypergeometric tau functions.}
The role of the hypergeometric functions of matrix argument in form of KP tau functions presented in \cite{OS-TMP}
was marked in \cite{Goulden-Jackson-2008} in the context of combinatorial problems.
Hypergeometric tau function of the two-component KP (2KP) may be written as
\be\label{hyperg2KP-sums}
\sum_{\lambda \atop \ell(\lambda)\le N} \, r_\lambda(n)\, c^{|\lambda|}\,s_\lambda(\bpow) \,
s_\lambda(\bbpow) \,
=
: \tau_r^{\rm 2KP}(N,n,\bpow,{\bbpow})
\ee
where $r_\lambda(n)$ is the same as in (\ref{hypergBKP-sums}).
 Here two independent sets $\bpow=(p_1,p_2,\dots)$ and $\bbpow=({\bar p}_1,{\bar p}_2,)$ and
 two discrete parameters $N$ and $n$ play the role of 2KP higher times. (We do not mark the dependence of the right hand
 side on the constant $c$ since it is trivial.)
 Then hypergeometric tau functions of
 2KP and BKP hierarchies are related:
 \be\label{2KP-to-BKP}
\left[ e^{\sum_{i>0}\frac {i}{2}\frac{\partial^2}{\partial {\bar p}_i^2}+
\sum_{i>0,\,{\rm odd}}\frac{\partial}{\partial {\bar p}_i}}\cdot
\tau_r^{\rm 2KP}(N,n,\bpow,{\bbpow})\right]_{\bbpow =0}\,=\,\tau_r^{\rm B}(N,n,\bpow)
 \ee
 which follows from
 (\ref{Laplace=sum-Schurs}) and (\ref{Schur-orthogonality}):
 \be
\left[ e^{\sum_{i>0}\frac {i}{2}\frac{\partial^2}{\partial p_i^2}+
\sum_{i>0,\,{\rm odd}}\frac{\partial}{\partial p_i}}\cdot s_\lambda(\bpow)\right]_{\bpow=0} =1,
\quad
 \ee

 \paragraph{Hypergeometric tau functions via the vertex operators  \cite{NO-LMP} .}
 
We chose two different types of parameterizations of the function $r$ which defines the content product
in (\ref{hypergBKP-sums}). The first is
\be\label{r-MirMorNat}
\qquad({\rm I})\qquad\qquad\quad
r(x)= \,\exp \sum_{m>0}\,\frac 1m \zeta_m h^m x^m\,
\qquad\qquad\qquad\qquad
\ee
The second is
\be\label{r-new-trig}
\,({\rm II})\qquad\qquad\quad
r(x)= \, \texttt{t}^{x \xi_0 }\exp \sum_{m\neq 1}\,\frac 1m \xi_m \texttt{t}^{mx}\,
\qquad\qquad\qquad
\ee
The complex number $\texttt{t}$ and sets $\{\zeta_m,\,m > 0\}$ and $\{\xi_m,\,m\in\mathbb{Z} \}$
are free parameters. Similar to \cite{Okounkov-2000},\cite{Harnad-2014} we introduce auxiliary 
parameters $c$ and $h$, the powers of $c$ count the degree of covering maps while the powers of the
parameter $\frac 1h$ which enters (\ref{r-MirMorNat}) count the Euler characteristic
of the covers.

Parametrizations (\ref{r-MirMorNat}) and (\ref{r-new-trig}) may be re-writen respectively as
\be\label{choice-r-1}
r(\zeta,h;x)=\,\exp V(\zeta,hx)
\ee
and
  \be\label{choice-r-2}
r(\xi,x|\texttt{t})=\,e^{V(\xi_+,\texttt{t}^x) +\xi_0 x\log \texttt{t} +V(\xi_-,\texttt{t}^{-x})}=\,
e^{\sum_{m\neq 0}\frac{1-\texttt{t}^m}{m\texttt{t}^m}\texttt{p}^*_m \texttt{t}^{mx}+\xi_0 x\log \texttt{t}}
\ee
where $\xi$ is the collection of parameters $\xi_0$ and $\xi_\pm=(\xi_{\pm 1},\xi_{\pm 2},\dots)$, and where
$V$ is defined by (\ref{V}).

 From bosonization formulae of \cite{JM},\cite{PogrebSushko}
 in \cite{OST-I} tau functions (\ref{hypergBKP-sums}) were presented in terms of an action of the vertex operators.
 For $r$ given by (\ref{r-new-trig}) (the same: by (\ref{choice-r-2})) the tau function (\ref{hypergBKP-sums}) may be written as
 \be\label{tau-via-vertex-t}
 \tau_r^B(N,n,\bpow) =\,\frac{1}{g(n)}\,
 e^{\xi_0 {\hat h}_2(n) \log\texttt{t} +\sum_{m\neq 0} \texttt{p}^*_m {\hat h}(n,\texttt{t}^m)}
 \cdot \sum_{\lambda\atop \ell(\lambda)\le N}\,c^{|\lambda|}s_\lambda(\bpow)
\ee
where ${\hat h}(n,\texttt{t}^m) \, (m\in\mathbb{Z})$ are commuting operators defined as vertex operators
\be\label{hat-h-t}
{\hat h}(n,\texttt{t})\, :=\, \texttt{t}^{n}\res_z \,\frac {dz}{z}
e^{\sum_{i>0} (\texttt{t}^i-1) \frac{z^i p_i}{i}   }\,
e^{-\sum_{i>0} (\texttt{t}^{-i}-1) z^{-i} \frac{\partial }{\partial p_i} }
\ee
and where ${\hat h}_2(n)$ is determined by the generating series
${\hat h}(n,e^{\epsilon})=: 1+ \sum_{i\ge 0} \frac{\epsilon^{i+1}}{(i+1)!} {\hat h}_i(n)$.
The operators ${\hat h}_i(n)$ were written down in \cite{AMMN-2011} in the most explicit way.
From (\ref{hat-h-t}) we get
\[
{\hat h}_0(n) = n\,,\qquad {\hat h}_1(n) = n^2 + \sum_{i>0} ip_i\frac{\partial}{\partial p_i}\,,
\]
\be
{\hat h}_2(n)= n^3 + \sum_{i,j}
\left((i+j)p_ip_j\frac{\partial}{\partial p_{i+j}} + ij p_{i+j}\frac{\partial^2}{\partial p_{i}\partial p_{j}}\right)
\ee
In particular the operator ${\hat h}_2(0)$ is known as the cut-and-join operator which was introduced in \cite{GJ}.

For $r$ given by (\ref{r-MirMorNat}) (the same: by (\ref{choice-r-1})) tau function (\ref{hypergBKP-sums}) may be written as
\[
 \tau_r^B(N,n,\bpow)=\,\frac{1}{g(n)}\,e^{\sum_{m>0} p^*_m {\hat h}_m(n) }
 \cdot \sum_{\lambda,\,\ell(\lambda)\le N}\,c^{|\lambda|} s_\lambda(\bpow)
\]

{\bf Example.} Let
\be\label{xi-p*}
V(\zeta,x)=V(\bpow^*,x-1)-V(\bpow^*,x)
\ee

For $N=+\infty$, $n=0$ and $r(x)=e^{\zeta_1 x}$ (that is $p^*_m=0,\,m>2$, see (\ref{xi-p*}))
 we get
\be\label{Example-vertex}
\tau^{\rm B}_r(\bpow) =\sum_{\lambda} e^{\zeta_1 \varphi_\lambda(\Gamma)}c^{|\lambda|} s_\lambda(\bpow)
=e^{\zeta_1 {\hat h}_2(0)}\cdot
e^{\sum_{m>0} \frac {c^2}{2m}p_m^2 +\frac{c}{2m-1}p_{2m-1}}
\ee

\subsection{Hypergeometric BKP tau function in terms of free Fermi fields \label{fermionic-appendix} \cite{OST-I}}
Details may be found in \cite{OS-2000, OST-I}.
Let
 $\{\psi_i$, $\psi_i^\dag$, $i \in \mathbb{Z}\}$ are Fermi creation and
annihilation operators that  satisfy the usual anticommutation relations and vacuum annihilation conditions
\[
[\psi_i, \ \psi_j]_+ = \delta_{i,j}, \quad \psi_i | n\rangle =
\psi_{-i-1} | n\rangle =0,\quad   i< n
 \]
In contrast to the DKP hierarchy introduced in \cite{JM} for the BKP hierarchy introduced in \cite{KvdLbispec}
one needs an additional Fermi mode  $\phi$ which anticommutes with each other
Fermi operator except itself: $\phi^2=\frac 12$, and
 $\phi|0\r=\frac{1}{\sqrt{2}}|0\r$, see \cite{KvdLbispec}. Then the hypergeometric BKP tau function introduced in
 \cite{OST-I} may be written as
 \[
 g(n)\tau^{\rm B}_r(N,n,\bpow) =
 \l n| e^{\sum_{m>0} \frac 1m J_m p_m}
 e^{\sum_{i < 0} U_i \psi_i^\dag \psi_i -\sum_{i \ge 0} U_i \psi_i\psi_i^\dag }
 e^{\sum_{i>j} \psi_i\psi_j\,-\sqrt{2}\,\phi \sum_{i\in\mathbb{Z}} \psi_i}|n-N\r =
 \]
 \be\label{hyper-via-U,r}
= \sum_{\lambda\atop \ell(\lambda)\le N} \,e^{-U_\lambda(n)} s_\lambda(\bpow)
 = g(n)
\sum_{\lambda\atop \ell(\lambda)\le N}  r_\lambda(n)s_\lambda(\bpow)
\ee
where  $J_m=\sum_{i\in\mathbb{Z}}\,\psi_i\psi^\dag_{i+m},\ m>0$,
$U_\lambda(n)=\sum_{i} U_{h_i+n}$,
$r(i)=e^{U_{i-1}-U_{i}}$
and
\bea
 e^{-U_0+\cdots -U_{n-1}}\quad {\rm if}\,\, n>0 \qquad\qquad\qquad \\
g(n)\,:=\,\l n|  e^{\sum_{i < 0} U_i \psi_i^\dag \psi_i -\sum_{i \ge 0} U_i \psi_i\psi_i^\dag }
|n\r\,= \qquad\qquad\qquad\qquad\qquad
1\quad {\rm if} \,\,n=0 \qquad\qquad\qquad
\label{g(n)}
\\
e^{U_{-1}+\cdots U_{n}}\quad {\rm if}\,\, n<0 \qquad\qquad\qquad
\eea
In (\ref{hyper-via-U,r}) the summation runs over all partitions whose length do not exceed $N$.
\br\label{DKPvsBKP}
Let us note that without the additional Fermi mode $\phi$ the summation range in (\ref{hyper-via-U,r}) does not
include partitions with odd partition lengths. One can avoid this restriction by introducing a pair of DKP tau
functions which seems unnatural.
\er
Apart of (\ref{hyper-via-U,r}) the same series without the restriction $\ell(\lambda)\le N$ is the example
of the BKP tau function however it is related to the single value $n=0$, the $n$-dependence destroys the
simple form of such tau function, see \cite{OST-I}.

\subsection{BKP tau functions generating Hurwitz numbers \label{BKP-tau-Hurwitz-section} \cite{NO-LMP}}

As we shall see the hypergeometric tau functions generate weighted sums of Hurwitz numbers.
However there exist special cases when one gets Hurwitz numbers themselves, this is based on Remark \ref{special-cases}.

We will make difference between the parameterizations I and II.

First, let us write down the simplest case of a single branch point related to all $r=1$ and $N=\infty$.
This case is described by $\tau_1^{\rm B}$  where it is reasonable to produce the change $p_m\to h^{-1}c^m p_m$.
We get
\be\label{single-branch-point}
e^{\frac {1}{h^2}\sum_{m>0} \frac {1}{2m}p_m^2 c^{2m} +\frac 1h\sum_{m {\rm odd}} \frac 1m p_m c^m}=
\sum_{d>0} c^d 
\sum_{\Delta\atop |\Delta|=d} h^{-\ell(\Delta)} \bpow_\Delta
H^{1,a}(d;\Delta)
\ee
where $a=0$ and if $\Delta=(1^d)$, and where $a=1$  and
 otherwise. Then $H^{1,1}(d;\Delta)$ is the Hurwitz number 
describing $d$-fold covering of $\mathbb{RP}^2$ with a single
branch point of type $\Delta=(d_1,\dots,d_l),\,|\Delta|=d$ by a (not necessarily connected) Klein surface of
Euler characteristic $\textsc{e}'=\ell(\Delta)$. For instance, for $d=3$, $\textsc{e}'=1$ we get
$H^{1,1}(3;\Delta)=\frac 13\delta_{\Delta,(3)}$.
For unbranched coverings (that is for $a=0$, $\textsc{e}'=d$) we get formula (\ref{unbranched}).

Next let us notice that the exponent of the left hand side of (\ref{single-branch-point}) is the generating series of the
connected Hurwitz numbers
\[
 \frac{1}{h^2}\sum_{d=2m}c^{2m} p_m^2 H_{\rm con}^{1,1}\left(d;(m,m)\right) +
 \frac{1}{h} \sum_{d=2m-1} c^{2m-1}p_{2m-1} H_{\rm con}^{1,1}\left(d;(2m-1)\right)
\]
where $H_{{\rm con}}^{1,1}$ describes $d$-fold covering either by the Riemann
sphere ($d=2m$) or by the projective plane ($d=2m-1$). These are the only ways to cover $\mathbb{RP}^2$
by a connected surface for the case of the single branch point.
The geometrical meaning of the exponent in (\ref{single-branch-point}) may be explained as follows. The projective plain may be viewed as the unit disk with the identification
of the opposite points $z$ and $-z$ on the boarder: $|z|=1$. In case we cover the Riemann sphere by the Riemann sphere $z\to z^m$ we get
two critical points with the same profiles. However we cover $\mathbb{RP}^2$ by the Riemann sphere, then we have the composition of the
mapping $z\to z^{m}$ on the
Riemann sphere and the factorization by antipodal involution $z\to - \frac{1}{\bar z}$. Thus we have the ramification profile $(m,m)$
at the single critical point $0$ of $\mathbb{RP}^2$.
The automorphism group is the dihedral group of the order $2m$ which consists of rotations on $\frac{2\pi }{m}$ and antipodal involution
$z\to -\frac{1}{\bar z}$.
Thus we get that $H_{\rm con}^{1,1}\left(d;(m,m)\right)=\frac{1}{2m}$ which is the factor in the first sum in the exponent in
(\ref{single-branch-point}). 
Now let us cover $\mathbb{RP}^2$ by $\mathbb{RP}^2$ via $z\to z^d$. For even $d$ we have the critical point
$0$, in addition each point of the unit
circle $|z|=1$ is critical (a folding), while from the beginning we restrict our consideration only on isolated critical points.
For odd $d=2m-1$ there is
the single critical point $0$, the automorphism group consists of rotations on the angle $\frac{2\pi}{2m-1}$. Thus in this case
$H^{1,1}\left(d;(2m-1)\right)=\frac{1}{2m-1}$ which is the factor in the second sum in the exponent in (\ref{single-branch-point}).

Next, let us consider BKP hypergeometric function in the parametrization I where we put
$\zeta_1=\beta +\sum_{i=1}^L a_i^{-1} $ and $\zeta_k=\sum_{i=1}^L a_i^{-k} $
\be\label{tau-Hurwitz-themselves}
H^{1,b+m+1}(d; \underbrace{\Gamma,\dots,\Gamma}_b,\underbrace{(d),\dots,(d)}_m,\Delta )= c^{-d}h^{\textsc{e}'}
\left[ \tau(N=+\infty,0,\bpow|\zeta) \right]_{b,m}
\ee
where the brackets $[*]_{b,m}$ means the picking up the factor of the term $\bpow_\Delta \prod_{i=1} a_i^{-1}$
which counts $d$-fold covers of $\mathbb{RP}^2$ with the following ramification type: there are $b$ simple branch points,
$m$ maximally ramified branch points, and one branch point of type $\Delta=(d_1,\dots,d_l)$. Each cover is a
connected Klein surface in case $m>0$ (and
not necessarily connected one in case $m=0$).
The Euler characterisitic of the cover is
$\textsc{e}'=\ell(\Delta)-b -m(d-1)$. 

Let us also recall that according to the correspondence (\ref{H-H}) 
from Proposition \ref{d-cycle-proposition} we conclude that the projective Hurwitz number of (\ref{tau-Hurwitz-themselves}) 
may be equated to Hurwitz numbers related to different base surfaces.

 \subsection{Hirota equations for the BKP tau function with two discrete time variables.\label{N-n-BKP-Hirota}}

 The BKP hierarchy we are interested in was introduced in \cite{KvdLbispec}.
 In  \cite{KvdLbispec} the BKP tau function $\tau^{\rm B}(N,\bpow)$ does not contain the discrete variable $n$.
 We need in a slightly general version of BKP hierarchy which includes $n$ as the higher time parameter,
 see \cite{OST-I} and \cite{LeurO-2014}.
 Hirota equations for the tau functions $\tau^{\rm B}(N,n,\bpow)$ of this modified BKP hierarchy read
\bea\label{Hirota-N-n-BKP-OST'}
  \oint\frac{dz}{2\pi i}z^{N'-N-1}e^{V(\bpow'-\bpow,z)}
  \tau(N'-1,n,\bpow'-[z^{-1}])
  \tau(N+1,n+1,\bpow+[z^{-1}]) \nonumber\\
+ \oint\frac{dz}{2\pi i}z^{N-N'-3}e^{V(\bpow-\bpow',z)}
  \tau(N'+1,n+2,\bpow'+[z^{-1}])
  \tau(N-1,n-1,\bpow-[z^{-1}]) \nonumber\\
=
  \tau(N'+1,n+1,\bpow')
  \tau(N-1,n,\bpow)
- \frac{1}{2}(1-(-1)^{N'+N})
  \tau(N',n+1,\bpow'|g)\tau(N,n,\bpow)
\eea
 and
\bea\label{Hirota-N-BKP(deLeur)'}
  \oint\frac{dz}{2\pi i}z^{N'-N-2}e^{V(\bpow'-\bpow,z)}
  \tau(N'-1,n-1,\bpow'-[z^{-1}])\tau(N+1,n+1,\bpow +[z^{-1}]) \nonumber\\
+ \oint\frac{dz}{2\pi i}z^{N-N'-2}e^{V(\bpow-\bpow',z)}
  \tau(N'+1,n+1,\bpow'+[z^{-1}])\tau(N-1,n-1,\bpow'-[z^{-1}]) \nonumber\\
= \frac{1}{2}(1-(-1)^{N'+N})\tau(N',n,\bpow')\tau(N,n,\bpow)
\eea
 Here $\bpow=(p_1,p_2,\dots)$, $\bpow'=(p'_1,p'_2,\dots)$.
The symbol $\bpow +[z^{-1}]$ denotes the set $\left( p_1+z^{-1},p_2+z^{-2}, p_3+z^{-3},\dots \right)$ and
$V$ is defined by (\ref{V}).

Equations (\ref{Hirota-N-BKP(deLeur)'}) are the same as in \cite{KvdLbispec} while equations
(\ref{Hirota-N-n-BKP-OST'}) relate tau functions with different discrete time $n$ and were written down
in \cite{OST-I} and \cite{LeurO-2014}.

 Taking $N'=N+1$ and all $p_i=p_i',\,i\neq 2$ in (\ref{Hirota-N-BKP(deLeur)'})
and picking up the terms linear in $p'_2-p_2$ we obtain (\ref{Hirota-elementary-1'}).
Taking $N'=N+1$ and all $p_i=p_i',\,i\neq 1$ in (\ref{Hirota-N-n-BKP-OST'})
and picking up the terms linear in $p'_1-p_1$ we obtain (\ref{Hirota-elementary-2'})

The relation of the BKP hierarchy to the two- and three-component KP hierarchy was established in \cite{LeurO-2014}.
(It is more correct to tell - with three-component hierarchy with "frozen" third component).

\section{Matrix integrals as generating functions of Hurwitz numbers from \cite{NO-2014},\cite{NO-LMP}
\label{Matrix-integrals}}

In case the base surface is $\mathbb{CP}^1$ the set of examples of matrix integrals generating Hurwitz numbers were studied in
works \cite{Chekhov-2014},\cite{MelloKochRamgoolam},\cite{AMMN-2014},\cite{ChekhovAmbjorn},\cite{KZ},\cite{ZL},\cite{Zog}.
One can show that the perturbation series in coupling constants of these integrals (Feynman graphs) may be related to TL
(KP and two-component KP) hypergeometric tau functions. 
It actually means that these series generate Hurwitz numbers with at most two arbitrary profiles
(An arbitray profile corresponds to a certain term in the perturbation series in the coupling constants which are higher times.
The TL and 2-KP hierarchies there are two independent sets of higher times
which yeilds two critical points for Hurwitz numbers).
Meanwhile other profiles are subjects of certain conditions, and the origin of the additional profiles is the choice of the function
$r$ in content product factors in
hypergeometric tau functions (\ref{hyperg2KP-sums}).

Here, very briefly, we will write down few generating series for the $\mathbb{RP}^2$ Hurwitz numbers.
These series may be not tau functions themselves but may be presented as integrals of tau functions of matrix argument.
(The matrix argument, which we denote by a capital letter, say $X$, means that the power sum variables $\bpow$ are specified
as $p_i=\tr X^i,\,i>0$. Then instead of
$s_\lambda(\bpow)$, $\tau(\bpow)$ we write $s_\lambda(X)$ and $\tau(X)$). If a matrix integral in examples below is a BKP tau
function then it generates Hurwitz numbers with a single arbitrary profile and all other are subjects of restrictions
identical to those in $\mathbb{CP}^1$ case mentioned above.
In all examples $V$ is given by (\ref{V}). We also recall that the limiting values of $\bpow(\texttt{q},\texttt{t})$
given by (\ref{p(t,q)}) may be $\bpow(a)=(a,a,\dots)$ and $\bpow_\infty=(1,0,0,\dots)$. We also recall that numbers
$H^{\textsc{e},\textsc{f}}(d;\dots)$ are Hurwitz numbers only in case $d\le N$, $N$ is the size of matrices.

For more details of the $\mathbb{RP}^2$ case see \cite{NO-2014}. New development in \cite{NO-2014} with respect to
the consideration in \cite{O-2004-New} is the usage of products of matrices. 
Here we shall consider a few examples. 
All examples include the simplest BKP tau function, of matrix argument $X$,
\cite{OST-I} defined by (compare to (\ref{sum-Schurs}))
\be\label{vac-tau-BKP'}
 \tau_1^{\rm B}(X)\,:=\,\sum_\lambda \,s_\lambda(X)\,=
 e^{\frac 12 \sum_{m>0} \frac 1m\left(\tr X^m\right)^2 + \sum_{m>0,{\rm odd}}\frac{1}{m}\tr X^m}
= \frac{\det^{\frac 12}\frac{1+X}{1-X} }{\det^{\frac 12}\left( \mathbb{I}_N \otimes \mathbb{I}_N - X\otimes X\right)}
\ee
as the part of the integration measure. Other integrands are the simplest KP tau functions
$\tau_1^{\rm KP}(X,\bpow):=e^{\tr V(X,\bpow)}$ where $V$ is defined by (\ref{V}) where the parameters
$\bpow$ may be called coupling constants. The perturbation series in coupling constants are expressed
as sums of products of the Schur functions over partitions and are similar to the series we considered in
the previous sections.

{\bf Example 1.} The projective analog of Okounkov's generating series for double Hurwitz series as a model of normal matrices.
From the equality
\[
\left({2\pi}{\zeta_1^{-1}} \right)^{\frac 12} e^{\frac{(n\zeta_0)^2}{2\zeta_1}} e^{\zeta_0 nc+ \frac12 \zeta_1 c^2}\,
=\,
 \int_{\mathbb{R}} e^{x_i n\zeta_0 +(cx_i- \frac12 x^2_i)\zeta_1} dx_i ,
\]
 in a similar way as was done in \cite{OShiota-2004} using $\varphi_\lambda(\Gamma)=\sum_{(i.j)\in\lambda}(j-i)$,
 one can derive
\[
 e^{n|\lambda|\zeta_0}e^{\zeta_1 \varphi_\lambda(\Gamma)}\delta_{\lambda,\mu}\,=\,\textsc{k} \,
 \int  s_\lambda(M) s_\mu(M^\dag) \det \left(MM^\dag\right)^{n\zeta_0}
 e^{-\frac12 \zeta_1\tr \left( \log \left( MM^\dag\right)\right)^2} dM
\]
where $\textsc{k}$ is unimportant multiplier, where $M$ is a normal matrix with eigenvalues $z_1,\dots,z_N$ and $\log |z_i|=x_i$,
and where
$dM=\,d_*U\,\prod_{i<j}|z_i-z_j|^2\prod_{i=1}^N d^2 z_i$. Then the $\mathbb{RP}^2$ analogue of Okounkov's generating series
may be presented as the following integral
(\cite{Okounkov-2000}) may be written
\be\label{Okounkov-tau-normal-matrices-BKP}
\sum_{\lambda\atop \ell(\lambda)\le N}e^{n|\lambda|\zeta_0 +
\zeta_1 \varphi_\lambda(\Gamma)}
s_\lambda(\bpow)=\textsc{k}
 \int  e^{V(M,\bpow)}
 e^{\zeta_0 n\tr \log \left(MM^\dag\right)-\frac12 \zeta_1\left( \tr\log \left( MM^\dag\right)\right)^2}
 \tau^{\rm B}_1(M^\dag) dM
\ee
Recall that in the work \cite{Okounkov-2000} there were studied Hurwitz numbers with an arbitrary number of simple branch points
and two arbitrary profiles. In our analog, describing the coverings of the projective plane, an arbitrary profile
only one, because, unlike the Toda lattice, the hierarchy of BKP has only one set of (continuous) higher times.

A similar representation of the Okounkov $\mathbb{CP}^1$   was earlier presented in
\cite{AlexandrovZabrodin-Okounkov}. The corresponding string equation related to double Hurwitz numbers
was written down in \cite{TakasakiHurwitz}.

Below we use the following notations
 \begin{itemize}
  \item $  d_*U $ is the normalized Haar measure on $\mathbb{\mathbb{U}}(N)$: $\int_{\mathbb{U}(N)}d_*U =1$

  \item $Z$ is a complex matrix
    $$
d\Omega(Z,Z^\dag)  =\,\pi^{-n^2}\,e^{-\tr \left(ZZ^\dag\right)}\,
\prod_{i,j=1}^N \,d \Re Z_{ij}d \Im Z_{ij}
  $$

  \item Let $M$ be a Hermitian matrix the measure is defined
   $$
   dM= \, \prod_{i\le j}
d\Re M_{ij} \prod_{i<j} d\Im M
  $$

 \end{itemize}

It is known \cite{Mac}
\be\label{s-s-N_lambda-1}
\int s_\lambda(Z)s_\mu(Z^\dag)\,d\Omega(Z,Z^\dag) = (N)_\lambda\delta_{\lambda,\mu}
\ee
where $(N)_\lambda:=\prod_{(i.j)\in\lambda}(N+j-i)$ is the Pochhammer symbol
related to $\lambda$. A similar relation
was used in \cite{O-Acta},\cite{HO-2006},\cite{O-2004-New},\cite{AMMN-2014},\cite{OShiota-2004}, for models of Hermitian, complex
and normal matrices.

By $\mathbb{I}_N$ we shall denote the $N\times N$ unit matrix.
We  recall that
$$ s_\lambda(\mathbb{I}_N)=(N)_\lambda s_\lambda(\bpow_\infty)\,,
\qquad s_\lambda(\bpow_\infty) = \frac{{\rm dim}\lambda}{d!},\quad d=|\lambda|$$.

{\bf Example 2. Three branch points.}
The generating function for $\mathbb{RP}^2$ Hurwitz numbers with three ramification points, having three
arbitrary profiles:
\be\label{3-points-integral}
 \sum_{\lambda,\,\ell(\lambda) \le N}
 \frac{s_\lambda(\bpow^{(1)}) s_\lambda(\Lambda) s_\lambda(\bpow^{(2)})}{\left( s_\lambda(\bpow_\infty) \right)^2}
\ee
\[
 = \,\int \,\tau^{\rm B}_1\left( Z_1 \Lambda Z_2 \right)  \,\prod_{i=1,2} \,
  e^{V(\tr Z^\dag_i,\,\bpow^{(i)})}\,d\Omega(Z_i,Z^\dag_i)
\]
If $\bpow^{(2)}=\bpow(\texttt{q},\texttt{t})$ with any given parameters $\texttt{q},\texttt{t}$, and $\Lambda=\mathbb{I}_N$
then (\ref{3-points-integral}) is the hypergeometric BKP tau function. 

{\bf Example 3. 'Projective' Hermitian two-matrix model}.
The following integral
\[
\int \tau^{\rm B}_1(c M_2)  e^{\tr V(M_1,\bpow)+\tr (M_1 M_2)}dM_1dM_2 =
\sum_{\lambda}\,c^{|\lambda|} (N)_\lambda  s_\lambda(\bpow)
\]
where $M_1,M_2$ are Hermitian matrices is an example of the hypergeometric BKP tau function.
\br\label{GD-monotone-paths}
Using results of \cite{GJ} we can show that it is a projective analogue of the generating function of the
so-called \textit{strictly monotonic Hurwitz numbers} introduced by Goulden and Jackson. In the
projective case these numbers counts paths on Cayley graph of the symmetric group whose initial point is a given partition
while the end point is not fixed: we take the weighted sum over all possible end points, say, $\Delta$,  the weight is
given by $\chi(\Delta)$ of Lemma \ref{Hurwitz-down-Lemma}.
\er

{\bf Example 4. Unitary matrices.} Generating series for projective Hurwitz numbers with arbitrary profiles
in $n$ branch points and restricted profiles in other points:
\be\label{multimatrix-unitary-RP2'}
\int e^{\tr (c U_1^\dag \dots U_{n+m}^\dag)}
\left(\prod_{i=n+1}^{n+m} \tau^{\rm B}_1(U_i) d_*U_i \right)
\left(\prod_{i=1}^{n} \tau^{\rm KP}_1(U_i,\bpow^{(i)}) d_*U_i \right)
=
\ee
\[
\sum_{d\ge 0}c^d \left( d! \right)^{1-m} \sum_{\lambda,\, |\lambda|=
d\atop \ell(\lambda)\le N}\, \left(\frac{{\rm dim}\lambda}{d!}  \right)^{2-m}
\left(\frac{s_\lambda(\mathbb{I}_N)}{{\rm dim} \lambda} \right)^{1-m-n}
\prod_{i=1}^n \frac{s_\lambda(\bpow^{(i)})}{{\rm dim} \lambda}
\]

Here $\bpow^{(i)}$ are parameters. This series generate certain linear combination of Hurwitz numbers for base surfaces
with Euler characteristic $2-m,\,m\ge 0$.
The integral (\ref{multimatrix-unitary-RP2'}) is a BKP tau function in case the parameters
are specialized as $\bpow^{(i)}=\bpow(\texttt{q}_i,\texttt{t}_i),\,i=2,\dots,n$ with any values of
$\texttt{q}_i,\texttt{t}_i$,
and if in addition $m=1$. In case $n=1$ this BKP tau function may be viewed as an analogue of the generating function of
the so-called non-connected Bousquet-Melou-Schaeffer numbers
(see Example 2.16 in \cite{KazarianLando}).
In case $n=m=1$ we obtain the following BKP tau function
\[
\int \tau^{\rm B}_1(U_2)  e^{\tr V(U_1,\bpow)+\tr (cU_1^\dag U_2^\dag)}d_*U_1d_*U_2 =
\sum_{\lambda\atop \ell(\lambda)\le N}\,c^{|\lambda|}\frac{s_\lambda(\bpow)}{(N)_\lambda}
\]
If we compare this series with ones used in \cite{Goulden-Paquet-Novak},\cite{Harnad-2014} we can see that it is a projective analogue of the generating function of the
so-called \textit{weakly monotonic Hurwitz numbers}. In the projective case it counts paths on Cayley graph whose initial point is a given partition
while instead of a fixed end point we consider the sum over the all possible end points $\Delta$, 
with a weight given by $\chi(\Delta)$ of Lemma \ref{Hurwitz-down-Lemma}.

{\bf Example 5. Integrals over complex matrices}.

A pair of examples. 
An analogue of Belyi curves generating function \cite{Zog},\cite{Chekhov-2014} is as follows
 (compare also to (\ref{GJ-Hurwitz-number})):
\be
\sum_{l=1}^N N^l\sum_{ \Delta^{(1)},\dots,\Delta^{(n+1)}\atop \ell(\Delta^{n+1})=l} c^d
H^{\textsc{e},n+1}(d;\Delta^{(1)},\dots,\Delta^{(n+1)})
\prod_{i=1}^{n} \bpow^{(i)}_{\Delta^{(i)}}
=\sum_{\lambda}c^{|\lambda|}\frac{(d!)^{m-2}(N)_\lambda}{({\rm dim}\lambda)^{m-2}}\,
\prod_{i=1}^{n}\frac{s_\lambda(\bpow^{(i)})}{s_\lambda(\bpow_\infty)}
\ee
\be
=\int e^{\tr (cZ_1^\dag \dots Z_{n+m}^\dag)}
\left(\prod_{i=n+1}^{n+m} \tau^{\rm B}_1(Z_i) d\Omega(Z_i,Z_i^\dag) \right)
\left(\prod_{i=1}^{n} \tau^{\rm KP}_1(Z_i,\bpow^{(i)}) d\Omega(Z_i,Z_i^\dag) \right)
\ee
where $\textsc{e}=2-m$ is the Euler characteristic of the base surface.

The series in the following example generates the projective Hurwitz numbers themselves where to get rid
of the factor $(N)_\lambda$ in the sum over partitions we use mixed integration over $\mathbb{U}(N)$ and over
complex matrices:
\be
\sum_{ \Delta^{(1)},\dots,\Delta^{(n)}}\, c^d\,
H^{1,n}(d;\Delta^{(1)},\dots,\Delta^{(n)})\,
\prod_{i=1}^{n} \bpow^{(i)}_{\Delta^{(i)}}\,
=\sum_{\lambda,\,\ell(\lambda)\le N}\,c^{|\lambda|} \frac{{\rm dim}\lambda}{d!}\,
\prod_{i=1}^{n}\frac{s_\lambda(\bpow^{(i)})}{s_\lambda(\bpow_\infty)}
\ee
\be
=\,\int \tau_1^{\rm KP}(c U^\dag Z_1^\dag \cdots Z_k^\dag,\bpow^{(n)})\tau_1^{\rm B}(U)d_*U \prod_{i=1}^{n-1}
\tau_1^{\rm KP}(Z_i,\bpow^{(i)}) d\Omega(Z_i,Z_i^\dag)
\ee
Here $Z,Z_i,\,i=1,\dots,n-1$ are complex $N\times N$ matrices and $U\in\mathbb{U}(N)$. As in the previous examples
one can specify all sets $\bpow^{(i)}=\bpow(\texttt{q}_i,\texttt{t}_i),\,i=1,\dots,n$ except a single one which in 
this case has the meaning of the BKP higher times.

\end{document}